\documentclass[pdflatex,sn-mathphys-num]{sn-jnl}

\usepackage{graphicx}%
\usepackage{multirow}%
\usepackage{amsmath,amssymb,amsfonts}%
\usepackage{amsthm}%
\usepackage{mathrsfs}%
\usepackage[title]{appendix}%
\usepackage{xcolor}%
\usepackage{textcomp}%
\usepackage{manyfoot}%
\usepackage{booktabs}%
\usepackage{algorithm}%
\usepackage{algorithmicx}%
\usepackage{algpseudocode}%
\usepackage{listings}%
\usepackage{subcaption}

\theoremstyle{thmstyleone}%
\theoremstyle{thmstyletwo}%

\theoremstyle{thmstylethree}%

\begin{document}

\title[Article Title]{Quantum Black Hole as a Harmonic Oscillator from the Perspective of the Minimum Uncertainty Approach}


\author*[1,2]{\fnm{Wilfredo} \sur{Yupanqui Carpio}}\email{w.yupanqui@fisica.uaz.edu.mx}

\author[2]{\fnm{Octavio} \sur{Obreg\'on}}\email{octavio@fisica.ugto.mx}

\affil*[1]{\orgdiv{Unidad Acad\'emica de F\'isica}, \orgname{Universidad Aut\'onoma de Zacatecas}, \orgaddress{\street{Calzada Solidaridad esquina con Paseo a la Bufa S/N}, \city{Zacatecas}, \postcode{98060}, \state{Zacatecas}, \country{M\'exico}}}

\affil[2]{\orgdiv{Departamento de F\'isica, \orgname{Divisi\'on de Ciencias e Ingenier\'ias, Universidad de Guanajuato}}, \orgaddress{\street{Loma del Bosque 103}, \city{Le\'on}, \postcode{37150}, \state{Guanajuato}, \country{M\'exico}}}



\abstract{Starting from the eigenvalue equation for the mass of a black hole derived by Mäkelä and Repo, we show that, by reparametrizing the radial coordinate and the wave function, it can be rewritten as the eigenvalue equation of a quantum harmonic oscillator. We then study the interior of a Schwarzschild black hole using two quantization approaches. In the standard quantization, the area and mass spectra are discrete, characterized by a quantum number \(n\), but the wave function is not square-integrable, limiting its physical interpretation. In contrast, a minimal-uncertainty quantization approach yields an area spectrum that grows as \(n^2\), and consequently the mass \(M\) also increases. In this framework, the wave function is finite and square-integrable, with convergence requiring that the deformation parameter \(\beta\) be regulated by a discrete quantum number \(m\). The wave function exhibits quantum tunneling connecting the black hole interior with both its exterior and a white hole region, effects that disappear in the limit \(\beta \to 0\). These results demonstrate how minimal-length effects both regularize the wave function and modify the semiclassical structure of the black hole.}




\maketitle

\section{Introduction}
\label{sec:Introduction}
A quantum black hole is a theoretical concept that arises from the formulation of a quantized Hamiltonian, which is defined within the framework of general relativity. This concept is central to the ongoing efforts to develop a theory of quantum gravity, which would reconcile the two fundamental theories of modern physics: quantum mechanics and general relativity.

Several approaches have been explored to investigate the interior of a black hole within the framework of quantum gravity. For instance, in Loop Quantum Gravity (LQG) \cite{thiemann2008modern}, which is one of the leading non-perturbative approaches to the quantization of gravity, numerous studies have been conducted on both the interior and the entire spacetime of black holes. In these studies, the Hamiltonian describing the black hole's interior, expressed in terms of Ashtekar-Barbero variables, is quantized using the so-called polymer quantization \cite{Abhay_Ashtekar_2003,PhysRevD.76.044016,PhysRevD.95.065026,PhysRevD.92.104029,FLORESGONZALEZ2013394,PhysRevLett.110.211301}. This quantization method effectively introduces a parameter that establishes a minimal scale in the model, which allows for the avoidance of the singularity and also introduces a bounce from a black hole to a white hole in the vacuum case \cite{Corichi_2016,PhysRevD.78.064040,MORALESTECOTL2021168401,PhysRevD.76.104030,https://doi.org/10.1155/2008/459290,PhysRevD.103.084038,Sobrinho_2023}. Polymer quantization also induces certain modifications in the algebra of the theory at the quantum level, which can be viewed as an effective modification of the classical algebra. Other methods for quantizing the interior of black holes can be found in the works \cite{Adeifeoba_2018,olmo2016classical,Husain_2005,PhysRevD.73.104009,jalalzadeh2012quantization,PhysRevD.57.2349,Almeida_2023,PhysRevD.60.024009,PhysRevLett.98.181301,PhysRevD.81.023528}.

When gravity is incorporated into quantum measurement processes, led to the generalization of Heisenberg's uncertainty relation (HUR). This modification to the HUR is known as the Generalized Uncertainty Principle (GUP), which implements a minimal uncertainty in the position by modifying the ordinary uncertainty relation of quantum mechanics to accommodate deformations at high energies, typically at the Planck scale \cite{PhysRevD.52.1108}. On the other hand, GUP can be understood as an alternative quantization procedure that imposes a modified commutator between the position and a generalized momentum (or between generalized position and momentum \cite{PhysRevD.52.1108}, depending on the chosen representation) \cite{PhysRevD.85.024016,PhysRevD.107.126009}, resulting in a minimum uncertainty in position or momentum \cite{BOSSO2018498}. In \cite{BIZET2023137636}, it is demonstrated that the GUP arises from the consideration of non-extensive entropies that depend only on the probabilities. GUP is also derived from different proposals: in \cite{G_Veneziano_1986}, the scattering of strings at ultra-high energies is considered to analyze the divergences of quantum gravity at the Planck scale; in \cite{MAGGIORE199365}, a gedanken experiment is proposed to measure the area of the apparent horizon of black holes in the context of quantum gravity; and \cite{SCARDIGLI199939} explores the idea that spacetime in the Planck region fluctuates, leading to the possibility of virtual micro-black holes affecting the measurement process. 

As mentioned, the effects of GUP are significant in systems with energies close to the Planck scale. Particularly relevant examples of such systems include the early universe and the interiors of black holes, where quantum gravity effects are expected to dominate \cite{ryan1972hamiltonian,PhysRevD.28.2960, ryan2015homogeneous, PhysRev.160.1113}. Therefore, quantum cosmology, the branch of physics that studies these systems, is the appropriate field where this modified quantization rule is expected to have a considerable impact. In this context, taking advantage of the fact that the interior of a Schwarzschild black hole is isometric to the Kantowski-Sachs cosmological model, for the first time, the quantization based on the minimal uncertainty approach has been applied to the minisuperspace variables that describe the dynamics inside the black hole \cite{Bosso_2020}. This implies a modification of the Wheeler-DeWitt equation, which governs the quantum cosmological model, thereby characterizing a modified dynamic of the solution.

Following this line of applying the minimal uncertainty approach to Quantum Cosmology, several works have been published. For example, in \cite{Bosso_2021}, the classical Hamiltonian of the Schwarzschild black hole interior is considered within the Ashtekar-Barbero connection formalism. Inspired by models based on the Generalized Uncertainty Principle, the canonical algebra of the model is deformed, leading to the derivation of the effective dynamics. This deformation results in the resolution of the black hole singularity by introducing a minimum nonzero radius for the infalling two-spheres. Recently, in \cite{MELCHOR2024116584}, starting from the proposal of a new reduced Hamiltonian, the classical black hole singularity is resolved by replacing it with an effective bounce that connects the interior of a black hole with the interior of a white hole. This bounce occurs in the region near the Planck scale, where a new event horizon emerges. Crossing this horizon changes the nature of the interval from spatial to temporal outside the white hole. Finally, in \cite{Bosso_2024}, the interior of a Schwarzschild black hole was quantized using the minimal uncertainty approach suitable for the Ashtekar-Barbero connection variables. As a result, it was found that all interior states remain well-defined and square-integrable. Moreover, the expectation value of the Kretschmann scalar remains finite throughout the entire interior region of the black hole, particularly in the area where the classical singularity used to reside, indicating the resolution of the black hole singularity. Additionally, a minimum value for the radius of the 2-spheres was also identified.

In this work, we focus on the eigenvalue equation derived in \cite{PhysRevD.57.4899} from a Hamiltonian that describes the spherically symmetric spacetime within the interior of the Reissner-Nordstr\"om black hole. The phase space characterizing this spacetime is parameterized by the charge \(Q\), the mass \(M\), and their respective conjugate momenta. Through a canonical transformation, configuration variables are obtained that naturally describe the dynamical properties of the black hole's interior. In \cite{obregon2001entropy}, the simplest case of this model is considered, taking \(Q=0\), and interestingly, by reparametrizing the black hole's radial coordinate and the wave function, the eigenvalue equation of a linear harmonic oscillator is obtained. As a result, it is found that the area spectrum, and therefore the Schwarzschild radius, is discrete and proportional to the square of the Planck length \cite{PhysRevD.57.4899,PhysRevD.9.3292}. 

Upon solving the eigenvalue equation for the black hole, modeled as a linear harmonic oscillator, and considering the semiclassical limit \( R_s \gg \ell_{\mathrm{Pl}} \), we find that the black hole singularity persists in the standard quantization scheme. To address this issue, we implement quantization using the minimal-uncertainty approach, which resolves the singularity by regularizing the wave function. Furthermore, this approach modifies the area spectrum and the black hole's radius, reflecting the effects of a minimal length scale.

Expressing black hole dynamics in a well-known form, such as that of the harmonic oscillator, provides significant advantages when studying some of its characteristics. Therefore, in this work, we focus on investigating the quantization of this black hole under two different quantization schemes: the standard scheme and the one based on minimum uncertainty. Special attention is given to the study of the singularity and the horizon within these two frameworks.

The paper is organized as follows: In section \ref{sec:Hamiltonian quantum theory of spherically symmetric black hole}, we provide a brief overview of the interior of the Schwarzschild black hole from the Hamiltonian perspective derived in \cite{PhysRevD.57.4899,PhysRevD.9.3292}. In section \ref{sec:Black hole as harmonic oscillator}, we apply an appropriate transformation to express the eigenvalue equation for the black hole mass as a quantum harmonic oscillator-like eigenvalue equation. Then, following the standard quantization procedure, i.e., using the usual commutation relation between canonical variables, we quantize the interior of the black hole. Section \ref{sec:Quantization employing the minimum uncertainty approach} is dedicated to the quantization of the black hole interior using the minimal uncertainty approach, in which the usual commutation relation is modified. We demonstrate that the resulting wave function is finite. Finally in section \ref{sec:Discussion and conclusion} we summarize our results and conclude. 

\section{Hamiltonian description of the black hole interior}
\label{sec:Hamiltonian quantum theory of spherically symmetric black hole}
The Reissner-Nordstr\"om metric is a static solution to the Einstein-Maxwell field equations, which corresponds to the gravitational field of a charged, non-rotating, spherically symmetric body of mass $M$. In spherical coordinates $(t,r,\theta,\varphi)$, this metric is
\begin{equation}
    ds^2 = -\left(1-\frac{R_s}{r}+\frac{R_Q^2}{r^2}\right)c^2dt^2 + \left(1-\frac{R_s}{r}+\frac{R_Q^2}{r^2}\right)^{-1}dr^2 + r^2d\Omega^2, 
\label{eq:Reissner_Nordstrom_metric}
\end{equation}
where $R_s=2GM/c^2$ is the  Schwarzschild radius and $R_Q^2=Q^2G/4\pi \epsilon_0c^4$ the characteristic length scale, in which $Q$ is the electric charge. In the limit that the charge $Q$ (or equivalently, the length scale $R_{Q}$) goes to zero, one recovers the Schwarzschild metric
\begin{equation}
ds^2=-\left(1-\frac{R_s}{r}\right)c^2dt^2+\left(1-\frac{R_s}{r}\right)^{-1}dr^2+r^2d\Omega^2,\label{eq:Schwarzschild_metric}
\end{equation}
with $r \in (0,\infty)$ being the radial coordinate. It is well-known that upon crossing the event horizon of the Schwarzschild black hole which is located at $R_s$, the timelike and spacelike curves switch their causal nature. This is because when an object crosses the event horizon of a black hole, its timelike trajectory inside the event horizon becomes spacelike, and it can no longer escape the gravitational attraction of the hole. So, one can obtain the interior metric by switching $t\leftrightarrow r$ in \eqref{eq:Schwarzschild_metric} \cite{doran2008interior}
\begin{equation}
ds^{2}=-\left(\frac{R_s}{t}-1\right)^{-1}dt^{2}+\left(\frac{R_s}{t}-1\right)dr^{2}+t^{2}d\Omega^2.\label{eq:sch-inter}
\end{equation}
Here $t$ is the Schwarzschild time coordinate which has a range $t\in(0,R_s)$. Consider the interior solution as measured by an observer at rest relatively to the space coordinates, i.e., $dr = d\Omega = 0$. In this case, from the metric \eqref{eq:sch-inter}, we have \cite{doran2008interior}
\begin{equation}
    \left(\frac{dt}{d\tau}\right)^2=\frac{R_s}{t}-1,\label{eq:Equanti_of_mot_t}
\end{equation}
where $\tau$ is the proper time.

In \cite{PhysRevD.54.2647, PhysRevD.57.4899}, a classical Hamiltonian corresponding to the metric in equation \eqref{eq:Reissner_Nordstrom_metric} is derived from the Einstein-Maxwell action. The constraint equations derived in this model allow us to define a reduced Hamiltonian of the Reissner-Nordstr\"om black hole in terms of the variables \(m\) and \(q\), which can be identified as the mass \(M\) and the charge \(Q\) of the hole. These variables, along with their conjugate canonical momenta \(p_m\) and \(p_q\), form a finite phase space. Suppose we choose an observer at the infinite asymptotic limit on the right side, at rest relative to the hole, and assume that the electric potential vanishes at this limit. In that case, the reduced Hamiltonian takes the form \(H = m\), which numerically corresponds to the mass \(M\) of the hole. Now, if the charge \(Q\) is fixed as an external parameter (in our case, we set \(Q = 0\)), it is possible to perform a canonical transformation from the phase space variables \((m, p_m)\) to the new canonical variables \((a, p_a)\), which naturally describe the dynamic properties inside the black hole. Consequently, the classical Hamiltonian, in terms of the variables \(a\) and \(p_a\), takes the form \cite{obregon2001entropy}
\begin{equation}
    H=\frac{p_a^2}{2a}+\frac{1}{2}a,\label{eq:Class_Hamilton}
\end{equation}
where the new coordinate $a$ and its conjugate momentum $p_a$ satisfy the classical algebra
\begin{equation}
    \{a,p_a\}=1.
\end{equation}
As noted in \cite{PhysRevD.57.4899}, the geometric interpretation of \( a \) is related to the radial coordinate of the hole. In our case, where an uncharged hole (\( R_Q = 0 \)) is considered, we conclude that \( 0 < a < R_s \). Therefore, we interpret the variable \( a \) as describing the classical dynamics within the interior of the Schwarzschild black hole.

The equation of motion for \( a \), in the case where \( Q = 0 \), is given by \cite{PhysRevD.54.4982}
\begin{equation}
    \dot{a}^2=\frac{R_s}{a}-1,\label{eq:Equa_motion_a}
\end{equation}
and by comparing this expression with \eqref{eq:Equanti_of_mot_t}, it becomes evident that the Schwarzschild time, \( t \), and the dynamical variable, \( a \), exhibit similar behavior. Thus, we can identify \(a\) as the radius of curvature of the two-sphere. It is also important to point out that this configuration variable is bounded along each classical trajectory, reaching its maximum value at the black hole's horizon. This implies that the spacetime dynamics, in terms of this variable, are confined to the interior of the hole. Consequently, an observer at infinity, outside of it, would perceive the exterior spacetime as static, which is physically reasonable \cite{PhysRevD.54.4982}.

Therefore, with these observations, we can express the interval \eqref{eq:sch-inter} in terms of the configuration variable \( a \)
\begin{equation}
ds^{2}=-\left(\frac{R_s}{a}-1\right)^{-1}da^{2}+\left(\frac{R_s}{a}-1\right)dr^{2}+a^{2}d\Omega^2.\label{eq:sch-inter_a}
\end{equation}
This justification is based on the fact that both variables \( t \) and \( a \) are confined to the interior of the hole, where $t\in(0,R_s)$ and $a\in(0,R_s)$. Furthermore, they describe the interior dynamics in the same manner, as can be verified by comparing equations \eqref{eq:Equanti_of_mot_t} and \eqref{eq:Equa_motion_a}.


\subsection{Interior quantization of the Schwarzschild black hole}
Since the Hamiltonian, \eqref{eq:Class_Hamilton}, describing the dynamic properties of the interior of the Schwarzschild spacetime is known, we can proceed with the canonical quantization of the spacetime in question. For this procedure, we consider a Hilbert space of the form \(L^2(\mathbb{R}^{+}, a^s \, \text{d}a)\), where \(s\) is a real number. In this space, the variable \(a\) acts as a multiplicative operator, while its conjugate momentum \(p_a\) is represented as a differential operator, \(p_a = -i \frac{\partial}{\partial a}\). In this representation, the corresponding symmetric Hamiltonian operator takes the form
\begin{equation}
    \hat{H}:=-\frac{1}{2}a^{-s-1}\frac{d}{da}\left(a^s\frac{d}{da}\right)+\frac{1}{2}a.\label{eq:Hamilt_OMT}
\end{equation}
As previously mentioned, the numerical value of the classical Hamiltonian for this model corresponds to the mass \(M\) of the Schwarzschild black hole \cite{PhysRevD.54.4982}. Therefore, we can consider an eigenvalue equation of the form
\begin{equation}
    \hat{H}\Psi(a)=M\Psi(a).\label{eq:Auto_val_equat}
\end{equation}
A fully dimensional form of the differential equation associated with \eqref{eq:Auto_val_equat} is derived in \cite{obregon2001entropy},
\begin{equation}
    \frac{\hbar^2G^2}{c^6}a^{-s-1}\frac{d}{da}\left(a^s\frac{d}{da}\right)\Psi(a)=\left(a-R_s\right)\Psi(a),\label{eq:Diff_equation_2}
\end{equation}
where $s$ is now an arbitrary factor ordering parameter. Equation \eqref{eq:Diff_equation_2} can be interpreted as analogous to a Wheeler-DeWitt equation, although the procedure by which it was derived is different. Therefore, its solution will yield a wave function representing a quantum black hole. Additionally, the phase space coordinates represented by the pair \((a, p_a)\) are not affected at the quantum level by their classical dependence on time \(t\) or radius \(r\). Thus, they can be interpreted as the variables of a minisuperspace.

Before solving equation \eqref{eq:Diff_equation_2}, we want to analyze some of its characteristics. As previously mentioned, the configuration variable \(a\) is classically confined within the black hole. Therefore, the term \(a - R_s\) on the right-hand side of \eqref{eq:Diff_equation_2} is negative. This implies, at least in the semiclassical approximation, an oscillatory behavior of the wave function \(\Psi(a)\) in the region \(0 < a < R_s\), and an exponential (decaying) behavior outside this region. This leads us to interpret our system as analogous to a particle confined in a potential well \cite{PhysRevD.57.4899}. Another observation is that the differential equation in \eqref{eq:Diff_equation_2} has a singular point at \(a = 0\), where the classical singularity resides. Therefore, it cannot evolve through this point.

Also, in \cite{PhysRevD.54.4982}, it has been shown that the Hamiltonian \( \hat{H} \) in \eqref{eq:Hamilt_OMT} is essentially self-adjoint for \( s \geq 4 \). On the other hand, for \( 1 \leq s < 4 \), the self-adjoint extensions of \( \hat{H} \) are characterized by a boundary condition at zero and are parametrized by \( U(1) \). It was also shown that the energy spectrum in \eqref{eq:Diff_equation_2} is discrete, bounded from below, meaning that the system has a ground state, and can be made positive. The lower bound of the spectrum implies that an infinite amount of energy cannot be extracted from the system, while the positivity of the spectrum aligns with the positive energy theorems of general relativity, which generally state that the ADM energy of the spacetime is always positive or zero when Einstein's field equations are satisfied. For \( s \geq 4 \), the energy (or mass) of the ground state is always positive, while for \( 1 \leq s < 4 \), it depends on the self-adjoint extension \cite{PhysRevD.57.4899}.

Defining \( a := x^{2/3} \) and \( \Psi(x) := x^{-\frac{1+2s}{6}} \chi(x) \) in equation \eqref{eq:Auto_val_equat}, together with \eqref{eq:Hamilt_OMT}, we obtain \cite{PhysRevD.54.4982}
\begin{equation}
    \frac{9}{8}\left[-\frac{d^2}{dx^2}+\frac{\nu(\nu-1)}{x^2}+\frac{4x^{2/3}}{9}\right]\chi(x)=M\chi(x),\label{eq:Diff_eq_in_x}
\end{equation}
where \( \nu = \frac{1 + 2s}{6} \) is defined. In this case, the new Hilbert space is given by \( L^2(\mathbb{R}^+; dx) \), as established in~\cite{PhysRevD.54.4982}.

We consider the case where the argument of \(\chi\) is small. To address this, we define a change of variable \( x = \left( \frac{9}{8M} \right)^{1/2} z \), and the eigenvalue equation \eqref{eq:Diff_eq_in_x} reduce to
\begin{equation}
    \left[-\frac{d^2}{dz^2}+\frac{\nu(\nu-1)}{z^2}-1+\left(\frac{3z}{8M^2}\right)^{2/3}\right]\chi(z)=0.\label{eq:eq:Diff_eq_in_z}
\end{equation}
The last term in this expression becomes asymptotically small for large values of \( M \), which also implies that \( x \) is asymptotically small. By omitting this last term, we obtain a differential equation of the form
\begin{equation}
    \left[-\frac{d^2}{dz^2}+\frac{\nu(\nu-1)}{z^2}-1\right]\chi(z)=0,\label{eq:eq:Diff_eq_in_z_1}
\end{equation}
from which, the solution is obtained
\begin{equation}
    \chi(x) = C_1 \sqrt{x} J_{\frac{1}{2} (2 \nu-1)}\left(\frac{(8M)^{1/2}x}{3}\right) + C_2 \sqrt{x} Y_{\frac{1}{2} (2 \nu-1)}\left(\frac{(8M)^{1/2}x}{3}\right), 
\label{eq:sol_diff_eq_z}
\end{equation}
where \( C_1 \) and \( C_2 \) are constants, and \( J_\mu(z) \) and \( Y_\mu(z) \) represent the Bessel functions of the first and second kinds, respectively. Here, we have expressed the solution in terms of the variable \( x \) instead of \( z \). The Bessel functions of the first kind, \( J_\mu(z) \), are finite at the origin, \( z = 0 \), for positive integer \( \mu \); whereas for negative non-integer \( \mu \), they diverge as \( x \) approaches zero. On the other hand, the Bessel functions of the second kind, \( Y_\mu(z) \), exhibit a singularity at the origin, \( z = 0 \). For this reason, we choose \( C_2 = 0 \) in the solution \eqref{eq:sol_diff_eq_z}, leaving us with
\begin{equation}
    \chi(x)=C_1 \sqrt{x} J_{\frac{1}{2} (2 \nu-1)}\left(\frac{(8M)^{1/2}x}{3}\right),\label{eq:Sol_diff_eq_x_finit}
\end{equation}
valid uniformly in any bounded region in $x$ \cite{PhysRevD.57.4899}. It should be noted that this solution was obtained for asymptotically large values of \( M \). Therefore, for large arguments, the Bessel function of the first kind behaves asymptotically as \( J_\mu(z) \sim \sqrt{\frac{2}{\pi z}} \cos\left(z - \frac{\mu \pi}{2} - \frac{\pi}{4}\right) \), which simplifies the eigenfunction \( \chi \) in \eqref{eq:Sol_diff_eq_x_finit} to
\begin{equation}
    \chi(x)\sim \cos\left(\frac{(8M)^{1/2}x}{3} - \frac{ \pi \nu}{2}\right).\label{eq:Sol_diff_eq_x_large_M}
\end{equation}
This expression is valid for the case \( \nu \geq \frac{3}{2} \), and also for \( \nu = \frac{1}{2} \) at large energies \cite{PhysRevD.57.4899}. In terms of the original variable \( a \), which represents the radius of the black hole, the wave function \( \Psi(a) \) takes the form
\begin{equation}
    \Psi(a)\sim \frac{1}{a^{3\nu/2}}\cos\left(\frac{(8M)^{1/2}a^{3/2}}{3} - \frac{ \pi \nu}{2}\right).\label{eq:Sol_diff_eq_a_large_M}
\end{equation}
Clearly, this wave function diverges at \( a = 0 \), the region where the physical singularity resides, but it exhibits oscillatory behavior in the interior region of the black hole, \( 0 < a < R_s \), the region of rapid oscillations at large $M$.

Since \(\chi\) decays exponentially in the region \(a > R_s\), the WKB approximation for the wave function in the rapidly oscillating region is given by \cite{PhysRevD.57.4899}
\begin{equation}
\chi_{\text{WKB}} \sim \cos \left[\frac{(8 M)^{1 / 2} x}{3}-\frac{\pi M^2}{2}+\frac{\pi}{4}+O\left(M^{-1 / 2}\right)\right].
\end{equation}
Comparing this with \eqref{eq:Sol_diff_eq_x_large_M} yields for the large eigenenergies the WKB estimate
\begin{equation}
M_{\text{WKB}}^2 \sim 2 n+\nu+\frac{1}{2}+o(1),\label{eq:Espect_area_MR}
\end{equation}
where $n$ is a large integer and $o(1)$ indicates a term that goes to zero at large $M$. It is concluded that, at the upper end of the spectrum, the asymptotic distribution of large eigenenergies yields an area spectrum for the black hole given by \eqref{eq:Espect_area_MR}. Bekenstein \cite{bekenstein1998quantum}, was the first to propose that the horizon area of a black hole is quantized in integer multiples of a fundamental unit, of the order of the Planck length squared, $l_{\text{Pl}}^2$. The quantized area spectrum takes the form $A = \alpha\, n\, l_{\text{Pl}}^2$, where $\alpha$ is a dimensionless constant of order unity, and $n$ is an integer. The implications of this area quantization for macroscopic physics were further developed by Bekenstein and Mukhanov~\cite{BEKENSTEIN19957}. To illustrate this, consider a Schwarzschild black hole. Its horizon area is related to the Schwarzschild mass $M$ by $A = 16\pi \left( \frac{l_{\text{Pl}}}{m_{\text{Pl}}} \right)^2 M^2$, where $m_{\text{Pl}}$ is the Planck mass. This relation implies that the mass $M$ can only take discrete values, as in equation~\eqref{eq:Espect_area_MR}. Consequently, during Hawking evaporation, the black hole can only undergo transitions between mass eigenstates corresponding to these discrete levels. As a result, the emitted radiation is composed of discrete quanta, with frequencies that are multiples of a fundamental frequency. This frequency is of the same order as the peak of Hawking’s blackbody spectrum, and its corresponding wavelength is comparable to the Schwarzschild radius of the black hole. Therefore, the radiation spectrum deviates from a continuous blackbody distribution in a way that is, in principle, macroscopically observable~\cite{PhysRevD.54.4982}.

\section{Black hole as harmonic oscillator}
\label{sec:Black hole as harmonic oscillator}
As shown in the previous section, the area spectrum of the black hole is quantized. 
This result was obtained by considering the asymptotic limit of large energies (masses), 
where the wave function---describing the eigenstate of the system---exhibits an oscillatory 
behavior within the black hole’s interior region. In this section, we revisit the same black hole, 
reformulating the corresponding eigenvalue equation in the form of a quantum harmonic oscillator.

It is interesting and somewhat surprising to note that the eigenvalue equation for a Schwarzschild black hole, as expressed in \eqref{eq:Diff_equation_2}, can be reparametrized as follows \cite{obregon2001entropy}
\begin{align}
    \Psi(a)=&\frac{U(a)}{a}, &x=&a-\frac{R_s}{2},\label{eq:Change}
\end{align}
 which leads to an eigenvalue equation resembling that of a linear harmonic oscillator
 \begin{equation}
     -\frac{\hbar^2G^2}{c^6}\left(\frac{d^2}{dx^2}+\frac{s-2}{x-R_s/2}\frac{d}{dx}-\frac{s-2}{(x-R_s/2)^2}\right)U(x)+x^2U(x)=\frac{R_s^2}{4}U(x),\label{eq:Diff_equation_3}
 \end{equation}
in particular, when the factor ordering parameter is set to \( s = 2 \), the expression in \eqref{eq:Diff_equation_3} reduces to a form analogous to the differential equation of a quantum harmonic oscillator
\begin{equation}
    \left(-\ell_{\mathrm{Pl}}^2\frac{d^2}{dx^2}+\frac{x^2}{\ell_{\mathrm{Pl}}^2}\right)U(x)=\frac{R_s^2}{4\ell_{\mathrm{Pl}}^2}U(x),\label{eq:Diff_equation_4}
\end{equation}
where $\ell_{\mathrm{Pl}}^2=\hbar G/c^3$ denotes the Planck length. It is important to note that the change of variables \( x = a - R_s/2 \), applied to the original domain \( a \in (0, \infty) \), results in a new variable \( x \in (-R_s/2, \infty) \). This domain does not cover the full real line and therefore lacks global translation invariance. As a consequence, standard Fourier analysis and the usual canonical representation of the pair \( (x, p_x) \) are not strictly justified in this setting. Now, therefore, the new Hilbert space will be \( L^2\left(\left(-\frac{R_s}{2}, \infty\right); \left(x + \frac{R_s}{2}\right)^2 dx\right) \).

For our purposes, we can define the momentum operator, conjugate to $x$, as a differential operator
\begin{equation}
    \hat{p}_x=-i\ell_{\mathrm{Pl}}\frac{d}{dx},
\end{equation}
which clearly corresponds to the differential operator on the left-hand side of equation \eqref{eq:Diff_equation_4}. Therefore, we can rewrite \eqref{eq:Diff_equation_4} as follows
\begin{equation}
    \left(\hat{p}_x^2+\frac{\hat{x}^2}{\ell_{\mathrm{Pl}}^2}\right)U(x)=\frac{R_s^2}{4\ell_{\mathrm{Pl}}^2}U(x).\label{eq:New_Hamilton}
\end{equation}
Our aim is to solve this eigenvalue equation in the momentum representation, where \( \hat{p}_x \) acts multiplicatively and \( \hat{x} \) as a differential operator. This choice is motivated by the fact that, in the following section, we will implement the quantization process based on the minimal uncertainty approach, where the representation of \(\hat{p}_x\) as a differential operator is not practically well-defined, at least to the best of our knowledge. Nevertheless, due to the restriction \( x \in \left(-\frac{R_s}{2}, \infty\right) \), the identification \( x = i \ell_{\mathrm{Pl}} \frac{d}{dp_x} \) cannot be freely applied. Nonetheless, in the semiclassical regime \( R_s \gg \ell_{\mathrm{Pl}} \) (for large values of $M$), the domain of \( x \) effectively approximates \( \mathbb{R} \). In this limit, we can define the Hilbert space as 
\( L^2\!\left(\mathbb{R}; \left(x + \tfrac{R_s}{2}\right)^2 dx \right) \), 
which naturally allows us to adopt the standard momentum-space representation
\begin{equation}
    \hat{p}_x = p_x, \qquad \hat{x} = i \ell_{\mathrm{Pl}} \frac{d}{dp_x},
\end{equation}
which satisfies the canonical commutation relation
\begin{equation}
    [\hat{x},\hat{p}_x]=i\ell_{\mathrm{Pl}}.\label{eq:Usual_conmut_rel_x_px}
\end{equation}
Within this approximation, the quantum dynamics is governed by
\begin{equation}
    \frac{d^2\bar{U}(p_x)}{dp_x^2}+(\kappa^2-p_x^2)\bar{U}(p_x)=0,\label{eq:Diff_equation_5}
\end{equation}
which corresponds to the Schrödinger equation of a harmonic oscillator in momentum space. Here we define $\kappa^2=R_s^2/4\ell_{\mathrm{Pl}}^2$. Clearly, this choice of the factor ordering parameter ensures that the Hamiltonian in \eqref{eq:Hamilt_OMT} remains self-adjoint. We emphasize that this representation should be understood as a \textit{formal approximation}, valid only in the limit \( R_s \to \infty \). A fully rigorous treatment would require defining the appropriate Hilbert space over the half-infinite domain \( x \in (-R_s/2, \infty) \) and analyzing the self-adjoint extensions of the relevant operators.

Equation \eqref{eq:Diff_equation_5} is known as Weber's differential equation, and its solution is given by
\begin{equation}
    \bar{U}_n(p_x)=N_ne^{-p_x^2/2} H_n(p_x),\label{eq:Usual_Solut_1}
\end{equation}
where $H_n(z)$ is the Hermite polynomials, $N_n$ a normalization constant
\begin{equation}
    N_n=\sqrt{\frac{1}{2^n\sqrt{\pi}n!}},\label{eq:Normalizat_const_Usuall}
\end{equation}
where \( n \) is a discrete quantum number taking values \( n = 0, 1, 2, \cdots \). According to this solution, the possible eigenvalues of the black hole horizon area are given by
\begin{equation}
    A_s(n)=32\pi\left(n+\frac{1}{2}\right)\ell_{\mathrm{Pl}}^2.\label{eq:Black_hol_Area_quantiz}
\end{equation}
This result has two important implications. First, it demonstrates that the area spectrum—and consequently, the energy (or mass) of the black hole—is bounded from below, ensuring positivity and, therefore, stability. Second, the spectrum given in \eqref{eq:Black_hol_Area_quantiz} is quantized and proportional to the square of the Planck length. On the other hand, given the quantization of the area \eqref{eq:Black_hol_Area_quantiz}, the Schwarzschild radius must also be quantized due to the relation $A_s(n)=4\pi R_s^2(n)$, and consequently the mass $M(n)$ of the black hole. Thus, from \eqref{eq:Black_hol_Area_quantiz} we obtain
\begin{equation}
    R_s(n)=2\sqrt{2\left(n+\frac{1}{2}\right)}\ell_{\mathrm{Pl}},\label{eq:Cuantiz_Schwar_radius}
\end{equation}
In our case, we are working in the semiclassical limit \( R_s \gg \ell_{\mathrm{Pl}} \), which in turn implies considering large values of the black hole mass \( M \). According to equation \eqref{eq:Cuantiz_Schwar_radius}, \( M \) is quantized and depends on the quantum number \( n \), which means that transitions between different states must also occur in discrete steps. Consequently, large values of \( M \) correspond to large values of \( n \). In this regime, the area spectrum in equation \eqref{eq:Black_hol_Area_quantiz} reduces to \( A_s(n) = 32\pi n \, \ell_{\mathrm{Pl}}^2 \), a result that agrees with those obtained in~\cite{bekenstein1998quantum,PhysRevD.57.4899,PhysRevD.54.4982,obregon2001entropy} within the WKB approximation. Moreover, the Schwarzschild radius in equation \eqref{eq:Cuantiz_Schwar_radius} asymptotically becomes \( R_s^2 = 8n \, \ell_{\mathrm{Pl}}^2 \). For higher energy states, the radius increases proportionally to the Planck length.

Although we have obtained the wave function for the interior of a Schwarzschild black hole under standard quantization in terms of the momentum variable \( p_x \), more direct physical insights about the hole can be gained from the configuration variable \( x \), since \( x \) is related to the black hole radius through \eqref{eq:Change}. Therefore, it is convenient to switch to the configuration-space representation by performing a standard Fourier transform of the form
\begin{equation}
    \phi(x)=\frac{1}{\sqrt{2\pi \ell_{\mathrm{Pl}}}}\int e^{ip_x x/\ell_{\mathrm{Pl}}}\ \bar{\phi}(p_x)\ dp_x.\label{eq:Usual_Fourie_trans}
\end{equation}
It should be noted that the range of the variable \( p_x \) is \( -\infty < p_x < \infty \). The Fourier transform of \eqref{eq:Usual_Solut_1} into configuration space can be performed using the exponential generating function of the Hermite polynomials. Then, from \eqref{eq:Usual_Fourie_trans}, one obtains
\begin{equation}
    U_n(x)=\frac{(i)^nN_n}{\sqrt{\ell_{\mathrm{Pl}}}}e^{-x^2/2\ell_{\mathrm{Pl}}^2} H_n(x/\ell_{\mathrm{Pl}}).\label{eq:Usual_Solut_x}
\end{equation}
In our setting, we are confined to large values of \( n \), and thus, for practical purposes, we may take the limit \( n \to \infty \). In this limit, the wave function \( \psi \) from equation \eqref{eq:Change}, rewritten in terms of \( x \) via equation \eqref{eq:Usual_Solut_x}, simplifies to
\begin{equation}
    \Psi_n(x)\sim\frac{1}{x+\frac{R_s(n)}{2}} \cos{\left[\frac{x\sqrt{2n}}{\ell_{\mathrm{Pl}}}-\frac{n\pi}{2}\right]},\label{eq:Usual_Solut_x_large_n}
\end{equation}
where the Hermite polynomials have been approximated using their asymptotic expansion in the limit \( n \to \infty \). The oscillatory behavior exhibited in equation \eqref{eq:Usual_Solut_x_large_n}, for large masses \(M\), is consistent with the result in \eqref{eq:Sol_diff_eq_a_large_M} and, similarly, has a singular point at \( x = -R_s/2=-\ell_{\mathrm{Pl}}\sqrt{2n} \). This solution rapidly decays for an observer at infinity (\( x \to \infty \)) in an asymptotically flat spacetime, ensuring that the black hole appears static from the exterior.

\begin{figure}[htbp]
    \centering

    \begin{subfigure}[t]{0.45\columnwidth}
        \centering
        \includegraphics[width=\linewidth]{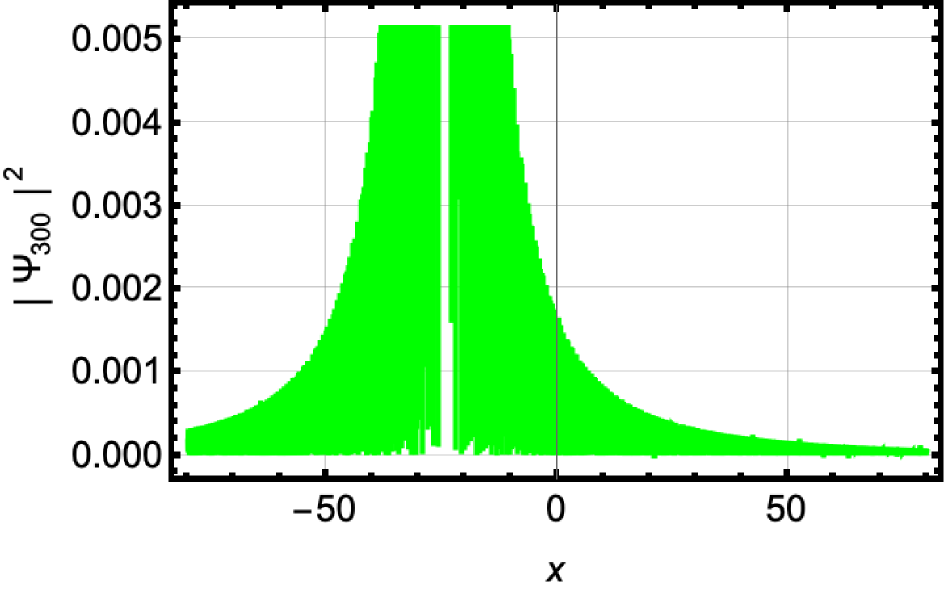}
        \subcaption{$n=300$}
        \label{fig:imagen1}
    \end{subfigure}
    \hfill
    \begin{subfigure}[t]{0.47\columnwidth}
        \centering
        \includegraphics[width=\linewidth]{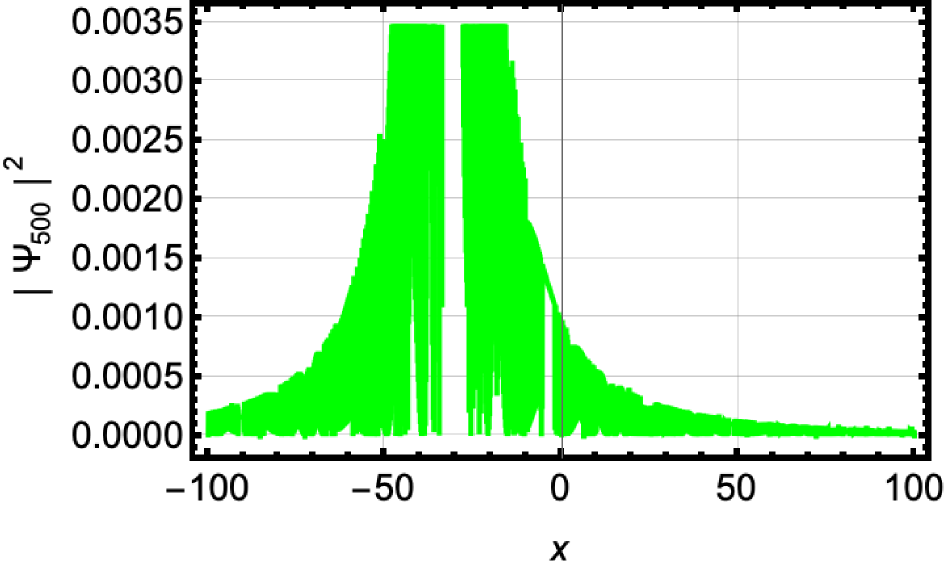}
        \subcaption{$n=500$}
        \label{fig:imagen2}
    \end{subfigure}

    \caption{Square of the wave function $\lvert \psi(x) \rvert^2$ plotted for different quantum numbers $n$.}
    \label{fig:Squar_wave_func_usuall}
\end{figure}

In Figure~\ref{fig:Squar_wave_func_usuall}, we present the squared wave function \eqref{eq:Usual_Solut_x_large_n} for large values of \(n\), plotted as a function of the variable \(x\). The probability density \( |\Psi_n(x)|^2 \) displays pronounced oscillations near the region \( x = -\ell_{\mathrm{Pl}}\sqrt{2n} \), where the wave function \eqref{eq:Usual_Solut_x_large_n} exhibits a singular point, and these oscillations gradually fade as \( x \to \infty \). Furthermore, as \(n\) increases, the number of oscillations also grows, in accordance with \eqref{eq:Sol_diff_eq_a_large_M} and \eqref{eq:Usual_Solut_x_large_n}. At sufficiently large distances from the black hole, these oscillations disappear altogether, implying that a distant observer would perceive the black hole as static.

We now need to determine whether the wave function derived from the standard quantization approach is square integrable. To this end, we integrate the squared modulus of \eqref{eq:Usual_Solut_x_large_n} over the semiclassical domain of \(x\), which yields
\begin{equation}
    \int_{-\infty}^\infty |\Psi_n(x)|^2\ \left(x + \frac{R_s}{2}\right)^2 dx \sim  \int_{-\infty}^{\infty}
    \cos^2{\left(\frac{x\sqrt{2n}}{\ell_{\mathrm{Pl}}}-\frac{n\pi}{2}\right)}\ dx \to\infty.
    \label{eq:Int_denom_K}
\end{equation}
We thus conclude that this wave function is not square integrable. As a result, the Hilbert space to which it would belong is not well defined, making it impossible to extract meaningful physical information. In particular, this prevents the computation of the expectation value of the Kretschmann scalar and obstructs any quantum description of the singularity inside the black hole, highlighting the limitations of the standard quantization approach in this context.

\section{Quantization employing the minimum uncertainty approach}
\label{sec:Quantization employing the minimum uncertainty approach}
As observed in the previous section, the standard quantization of the Schwarzschild black hole, in the semiclassical limit \( R_s \gg \ell_{\mathrm{Pl}} \), encounters serious difficulties: the resulting wave function diverges and is therefore not square integrable. In this section, we adopt a quantization scheme based on the minimum uncertainty approach, which incorporates a fundamental minimal length scale, typically of the order of the Planck length.

The Heisenberg uncertainty principle is a fundamental concept in quantum mechanics that states that certain pairs of physical properties of a particle cannot be precisely measured simultaneously. In those measurement processes, the gravitational interaction between particles was completely neglected, although this was somehow justified by the considerable weakness of gravity when compared with other fundamental interactions. However, gravity should be considered when elementary measurement processes are discussed in order to address fundamental questions in nature. Incorporating gravity into these quantum measurement processes led to the generalization of Heisenberg’s uncertainty relation.

For example, in string theory, when the gravitational interactions between a given pair of elementary particles are considered, the gravitational coupling constant $G_N$ (a constant characterizing the gravitational attraction between elementary particles) grows large at high energy and gives a nonrenormalizable perturbation theory (UV problem of quantum gravity). To solve this problem it is required that at the Planck length, the graviton and other particles turn out to be not points but one-dimensional objects, loops of string. As a result, is found that it is not possible to test distances shorter than the characteristic string length. This fact suggests a modification of the uncertainty relation at the Planck scale, and consequently the existence of a minimal observable length of the order of string size.

One way to incorporate the quantum effects of gravity into the measurement process is by 
modifying the standard Heisenberg uncertainty relation as follows~\cite{PhysRevD.52.1108}
\begin{eqnarray}
    \Delta q\Delta p\geq\frac{\hbar}{2}\left(1+\beta\left(\Delta p\right)^2\right),\label{Eqq.GUP.1}
\end{eqnarray}
where $\beta$ is know as the deformed parameter. In ordinary quantum mechanics, $\Delta q$ can be made arbitrarily small by letting $\Delta p$ grow correspondingly, this is no longer the case when \eqref{Eqq.GUP.1} is considered. If for decreasing $\Delta q$, $\Delta p$ increases, the new term $\beta\left(\Delta p\right)^2$ on the right-hand side of \eqref{Eqq.GUP.1} will eventually grow faster than the left-hand side. Hence $\Delta q$ can no longer be made arbitrarily small, searching a minimal uncertainty of the order $\hbar\sqrt{\beta}$. 
It allows one to express the idea that a minimal length $l_{\text{min}}$ should quantum theoretically be described as a minimal uncertainty in position measurements. 

This type of modified uncertainty relation, commonly referred to in the literature as the 
\textit{Generalized Uncertainty Principle} (GUP), has been derived in~\cite{BIZET2023137636} 
from considerations of non-extensive entropies. The GUP also emerges from a variety of other 
theoretical proposals. For instance, in~\cite{G_Veneziano_1986}, the scattering of strings 
at ultra-high energies is analyzed to address the divergences of quantum gravity at the 
Planck scale. In~\cite{MAGGIORE199365}, a \textit{gedanken} experiment is formulated to measure 
the area of the apparent horizon of black holes within a quantum-gravity framework. 
Furthermore,~\cite{SCARDIGLI199939} explores the notion that spacetime undergoes fluctuations 
at the Planck scale, potentially giving rise to virtual micro-black holes that affect the 
measurement process.

The modified commutation relation for the $\hat{q}$ and $\hat{p}$ operators associated with (\ref{Eqq.GUP.1}) is expressed as \cite{PhysRevD.52.1108}
\begin{eqnarray}
\left[\hat{q},\hat{p}\right]=i\hbar\left(1+\beta\  \hat{p}^2\right).\label{Eqq.GCR.1}
\end{eqnarray}
This deformation encodes the existence of a minimal length scale, typically of the 
order of the Planck length, as motivated in several approaches to quantum gravity. 
Therefore, $\beta$ plays the role of a deformation parameter that controls the strength 
of quantum–gravitational corrections. In the limit $\beta \to 0$, the standard Heisenberg 
algebra and the corresponding spectrum are recovered, while for finite values of $\beta$ 
the resulting modifications reflect the influence of minimal–uncertainty effects on the 
quantum dynamics of the system.

Due to the deformed commutator \eqref{Eqq.GCR.1} the operators $\hat{q}$ and $\hat{p}$ are not conjugates anymore. Now, these fundamental variables are to be high energy operators valid, in particular, at or near the Planck scale. They have non-linear representations, $\hat{q}=q\left(\hat{q}_0\right)$, $\hat{p}=p\left(\hat{p}_0\right)$ in terms of the variables $\hat{q}_0$, $\hat{p}_0$ which are position and momentum operators at low energies, obeying the standard Heisenberg algebra $\left[\hat{q}_0,\hat{p}_0\right]=i\hbar$.

This minimal uncertainty approach has yielded intriguing results when considering the quantum effects of gravity. For instance, we can cite some works \cite{Bosso_2020,Bosso_2021}. Motivated by these successes, we will apply this quantization approach in this section to resolve the singularity of the black hole under consideration. To implement the minimal uncertainty approach in the quantization procedure, the algebra \eqref{eq:Usual_conmut_rel_x_px} will be modified according to \eqref{Eqq.GCR.1} in order to achieve
\begin{eqnarray}
\left[\hat{x},\hat{p}_x\right]=il_{Pl}\left(1+\beta\  \hat{p}_x^2\right),\label{eq:Modf_conmut_x_px}
\end{eqnarray}
from this one can find the generalized uncertainty relation
\begin{equation}
    \Delta x\Delta p_x\geq\frac{l_{Pl}}{2}\left[1+\beta(\Delta p_x)^2\right],
\end{equation}
which corresponds to minimal uncertainty in $x$ of the order $l_{Pl}\sqrt{\beta}$. Therefore, $\beta$ effectively defines the magnitude of the minimal uncertainty effects.

To simplify our calculations and proceed with a standard quantization procedure, it is convenient to introduce a new variable, $p_{x_0}$, conjugate to $x$, that satisfies the usual commutation relation \cite{PhysRevD.85.024016,PhysRevD.107.126009}
\begin{align}
        [\hat{x},\hat{p}_{x_0}]=il_{Pl}.\label{eq:Usual_conmut_rela_x_p0}
\end{align}
For simplicity in the calculations, we will work in the representation where \(\hat{p}_{x_0}\) acts as a multiplicative operator and the position operator is represented as a differential operator \(\hat{x} = i l_{Pl} \, \partial / \partial p_{x_0}\). As in the preceding section, we assume that this holds only in the semiclassical limit \( R_s \gg \ell_{\mathrm{Pl}} \). The reason for choosing this representation is that it is not possible to clearly define the momentum operator \(\hat{p}_x\) as a differential operator that satisfies the commutation relation \eqref{eq:Modf_conmut_x_px} \cite{PhysRevD.52.1108}. Therefore, from \eqref{eq:Modf_conmut_x_px}, we can find the relation between the physical variable $p_x$ and the auxiliary $p_{x_0}$ by
\begin{equation}
    p_x=\frac{1}{\sqrt{\beta}}\tan\left(\sqrt{\beta }p_{x_0}\right),\label{eq:Self_adjoint_representation}
\end{equation}
which satisfies the modified commutation relation \eqref{eq:Modf_conmut_x_px}. Also, the domain of $p_{x_0}$ is restricted to $-\pi/2\sqrt{\beta}<p_{x_0}<\pi/2\sqrt{\beta}$ \cite{PhysRevD.85.024016}.

In this approach, the differential equation \eqref{eq:Diff_equation_5} is expressed in terms of the new canonically conjugate variables $x$ and $p_{x_0}$. Thus, in the $p_{x_0}$ space, it reads
\begin{equation}
    \frac{d^2\bar{U}(p_{x_0})}{dp_{x_0}^{2}}+\left(\kappa^2-\frac{\tan^2{(\sqrt{\beta}p_{x_0})}}{\beta}\right)\bar{U}(p_{x_0})=0.\label{eq:Diff_equation_GUP1}
\end{equation}
Using the new variable $\xi$ by the change $\xi=\sqrt{\beta}p_{x_0}$, the above equation can be written as
\begin{equation}
    \frac{d^2\bar{U}(\xi)}{d\xi^{2}}+\left(\epsilon-\frac{\tan^2{\xi}}{\beta^2}\right)\bar{U}(\xi)=0,\label{eq:Diff_equation_GUP2}
\end{equation}
here we denoted $\epsilon=\kappa^2/\beta=R_s^2/4\beta l_{Pl}^2$. This differential equation reduces to the hypergeometric one through the transformation $z=\sin^2 \xi$ and $y=\bar{U}\cos^m{\xi}$ \cite{zaitsev2002handbook}
\begin{equation}
    z(z-1)y''(z)+\left[(1-m)z-\frac{1}{2}\right]y'(z)-\frac{1}{4}(m+\epsilon)y(z)=0,\label{eq:Diff_equat_Hyperg}
\end{equation}
where $m$ is a root of the quadratic equation $m^2+m-\frac{1}{\beta^2}=0$, from which one has
\begin{equation}
    m=\frac{-\beta\pm\sqrt{4+\beta^2}}{2\beta}.\label{eq:m}
\end{equation}
Therefore, the general solution of \eqref{eq:Diff_equat_Hyperg} is given by
\begin{multline}
    y(z) = C_1 \, _2F_1\Bigg(
    \frac{\beta + \sqrt{\beta^2+4} + 2 \sqrt{\epsilon  \beta^2+1}}{4 \beta },
    \frac{\beta + \sqrt{\beta^2+4} - 2 \sqrt{\epsilon  \beta^2+1}}{4 \beta };
    \frac{1}{2}; z
    \Bigg) \\
    + C_2 z^{1/2} \, _2F_1\Bigg(
    \frac{\beta + \sqrt{\beta^2+4} + 2 \sqrt{\epsilon  \beta^2+1}}{4 \beta } + \frac{1}{2}, 
    \frac{\beta + \sqrt{\beta^2+4} - 2 \sqrt{\epsilon  \beta^2+1}}{4 \beta } + \frac{1}{2};
    \frac{3}{2}; z
    \Bigg).
    \label{eq:Wave_funct_GUP_px}
\end{multline}
Here, \( C_1 \) and \( C_2 \) are integration constants. The Gauss hypergeometric function \({}_2F_1(A,B;C;z)\) is defined as a series that converges for 
\(|z|<1\). At the boundary \(|z|=1\), convergence occurs if \(C-A-B>0\), divergence occurs if 
\(C-A-B\leq -1\), and the series is conditionally convergent when \(-1<C-A-B\leq 0\). Furthermore, the series reduces to a polynomial of degree \(n\) in \(z\) 
whenever \(A=-n\) or \(B=-n\) (\(n=0,1,2,\dots\)). The series is not defined when \(C=-m\) 
(\(m=0,1,2,\dots\)), unless one of the parameters \(A\) or \(B\) is a negative integer smaller than \(m\)~\cite{abramowitz1968handbook}.
In our case, from~\eqref{eq:Wave_funct_GUP_px}, yields for both hypergeometric functions $C-A-B = -\frac{\sqrt{\beta^2+4}}{2\beta}$. Convergence at the boundary would require \(\beta<0\), so that \(C-A-B>0\), but such a choice 
leads to unphysical results. The only consistent option is \(\beta>0\), which implies divergence of the 
series at \(|z|=1\). Therefore, square–integrability is guaranteed only when the series truncates, 
namely when either \(A\) or \(B\) equals a negative integer. This truncation condition ensures 
polynomial solutions, which are finite on the interval \(0 \leq z \leq 1\), and gives rise to the 
discrete spectrum discussed below.

For example, if we consider the hypergeometric function appearing in the first line of equation \eqref{eq:Wave_funct_GUP_px}, this condition for the first argument results in
\begin{equation}
    \frac{\beta +\sqrt{\beta ^2+4}+2 \sqrt{\epsilon  \beta ^2+1}}{4 \beta }=-n, \label{eq:Converg_condit_1_GUP}
\end{equation}
with \( n \) a positive integer. From this condition, one can derive the eigenvalues of the black hole's area by solving for \(\epsilon\) from equation \eqref{eq:Converg_condit_1_GUP}, considering that \(\epsilon = \frac{R_s^2}{4 \beta l_{\text{Pl}}^2}\) and \(A_s = 4 \pi R_s^2\). That is
\begin{equation}
     A_s^{\text{GUP}}(2n) = 32\pi\left(2n+\frac{1}{2}\right)l_{Pl}^2
    \left(\sqrt{1+\frac{\beta^2}{4}}+\frac{\beta}{2}\right)+ 16 (2n)^2\pi \beta l_{Pl}^2.
    \label{eq:Area_gup_correct_1}
\end{equation}
If, on the other hand, we apply the convergence condition to the second argument in $\left._2 F_1(A,-n;C;z)\right.$, we obtain exactly \eqref{eq:Area_gup_correct_1}, thus ensuring the convergence of the first term in the wave function \eqref{eq:Wave_funct_GUP_px}. When the same convergence criterion is applied to the hypergeometric function appearing in the second line of equation \eqref{eq:Wave_funct_GUP_px}, we find that the spectrum of the area, in this case, is
\begin{equation}
    A_s^{\text{GUP}}(2n+1) = 32\pi\left((2n+1)+\frac{1}{2}\right)l_{Pl}^2
    \left(\sqrt{1+\frac{\beta^2}{4}}+\frac{\beta}{2}\right)+ 16 (2n+1)^2\pi \beta l_{Pl}^2.\label{eq:Area_gup_correct_1_1}
\end{equation}
Combining both expressions, \eqref{eq:Area_gup_correct_1} and \eqref{eq:Area_gup_correct_1_1}, we obtain the discrete spectrum of the black hole's area
\begin{equation}
    A_s^{\text{GUP}}(n)=32\pi\left(n+\frac{1}{2}\right)l_{Pl}^2\left(\sqrt{1+\frac{\beta^2}{4}}+\frac{\beta}{2}\right)+16 n^2\pi \beta l_{Pl}^2.\label{eq:Area_gup_correct_2}
\end{equation}
It can be readily seen from this expression that, in the limit \( \beta \to 0 \), the usual area spectrum in~\eqref{eq:Black_hol_Area_quantiz} is recovered. In this case, the area spectrum grows as \( n^2 \), which differs from the behavior found in equations~\eqref{eq:Black_hol_Area_quantiz} and~\eqref{eq:Espect_area_MR}, where the area spectrum increases linearly with \( n \). In the semiclassical limit, corresponding to \( n \to \infty \), the effective area spectrum approaches \( A_s \sim 16\pi \beta n^2 \ell_{\mathrm{Pl}}^2 \). In the present case, the black hole horizon grows more rapidly with each transition to higher-energy (mass) states compared to the standard case. The same applies to the mass \( M \), which in this modified scenario still takes discrete values for large quantum states of the black hole.

It is worth noting that the quadratic dependence of the spectrum on the quantum number $n$ is a clear indication of an underlying $SU(1,1)$ algebraic structure. In unitary representations of $SU(1,1)$, the eigenvalues of the generator $K_0$ grow linearly with $n$, while the Casimir operator produces a quadratic dependence in $n$, which matches the structure found in our spectrum. This observation suggests that the Hamiltonian could, in principle, be rewritten in terms of the $SU(1,1)$ generators, and that the spectrum could be generated algebraically by ladder operators $K_\pm$. A full exploration of this algebraic approach lies beyond the scope of the present work, but it provides a promising direction for future investigation.


According to these convergence conditions, \eqref{eq:Converg_condit_1_GUP}, the wave function \eqref{eq:Wave_funct_GUP_px} is written, in terms of the original variable $\xi$, as
\begin{equation}
    y_n(\xi)=C_1\, _2F_1\left(-n,n+\alpha;\frac{1}{2};\sin^2{\xi}\right)+ C_2\sin{\xi} \, _2F_1\left(-n,n+\alpha+1;\frac{3}{2};\sin^2{\xi}\right),\label{eq:Wave_funct_GUP_px_2}
\end{equation}
where we have denoted
\begin{equation}
    \alpha=\frac{\beta+\sqrt{4+\beta^2}}{2\beta}.\label{eq:alpha_vs_beta}
\end{equation}
Here, if the negative sign is chosen instead of the positive in \eqref{eq:m}, we find that $-m=\alpha$. Additionally, considering the special double-$n$ formulas of the Gegenbauer polynomials \cite{frank2010nist}
\begin{equation}
    C_{2 n}^{(\alpha )}(\sin{\xi})=(-1)^n\binom{n+\alpha-1}{n}\, _2F_1\left(-n,n+\alpha;\frac{1}{2};\sin^2{\xi}\right),
\end{equation}
and
\begin{equation}
    C_{2 n+1}^{(\alpha )}(\sin{\xi})=2(-1)^n\alpha\binom{n+\alpha}{n}\sin{\xi}\, _2F_1\left(-n,n+\alpha+1;\frac{3}{2};\sin^2{\xi}\right),
\end{equation}
we can write the solution \eqref{eq:Wave_funct_GUP_px_2} in a compact form
\begin{equation}
    \bar{U}_n(p_{0_x})=\mathcal{N}_n C_{n}^{(\alpha )}(\sin{(\sqrt{\beta}p_{0_x}}))\cos^\alpha{(\sqrt{\beta}p_{0_x})}, \label{eq:Wave_funct_GUP_px_3}
\end{equation}
where the substitution $\xi=\sqrt{\beta}p_{x_0}$ has been made, and $\mathcal{N}_n$ is a constant that can be determined from the normalization condition of the wave function \eqref{eq:Wave_funct_GUP_px_3}. This is achieved by applying the normalization conditions of the Gegenbauer polynomials~\cite{abramowitz1968handbook}
\begin{equation}
    \int_{-1}^{1}(1-h^2)^{\alpha-1/2}[C_{n}^{(\alpha )}(h)]^2\ dh=\frac{2^{1-2\alpha}\pi\Gamma(n+2\alpha)}{(n+\alpha)\Gamma^2(\alpha)\Gamma(n+1)}.
\end{equation}
Then, we have
\begin{equation}
    \mathcal{N}_n=\left[\frac{\sqrt{\beta}(n+\alpha)\Gamma^2(\alpha)\Gamma(n+1)}{2^{1-2\alpha}\pi\Gamma(n+2\alpha)}\right]^{1/2}.
\end{equation}
Unlike the constant in \eqref{eq:Normalizat_const_Usuall}, this integration constant is expressed not only in terms of the quantum number $n$ but also in terms of the deformation parameter $\beta$, as shown in \eqref{eq:alpha_vs_beta}.

\subsection{Limit $\beta\to 0$}
\label{subsec:Limit beta to 0}
In the standard limit, where $\beta\to0$, the usual solution \eqref{eq:Usual_Solut_1} should be recovered. To verify this, consider from \eqref{eq:alpha_vs_beta} that for small values of the deformation parameter $\beta$, we have $\alpha=1/\beta$, which implies that when $\beta\to0$, $\alpha\to\infty$. On the other hand, from the relationship between the Gegenbauer polynomial $C_{n}^{(\alpha )}$ and the relativistic Hermite polynomial $H_{n}^{(\alpha )}$ \cite{10.1063/1.530606}, given by
\begin{equation}
    H_{n}^{(\alpha )}(\sqrt{\alpha}u)=\frac{n!}{\alpha^{n/2}}(1+u^2)^{n/2}C_{n}^{(\alpha )}\left(\frac{u}{\sqrt{1+u^2}}\right),
\end{equation}
we can express \eqref{eq:Wave_funct_GUP_px_3} as follows
\begin{equation}
    \bar{U}_n(p_{0_x})=\left[\frac{\sqrt{\beta}(n+\alpha)\Gamma^2(\alpha)\alpha^{n}}{2^{1-2\alpha}\pi\Gamma(n+2\alpha)n!}\right]^{1/2}\cos^{n+\alpha}{(\sqrt{\beta}p_{0_x})}H_{n}^{(\alpha )}(\sqrt{\alpha}\tan{(\sqrt{\beta}p_{0_x})}),\label{eq:Wave_funct_GUP_px_4}
\end{equation}
where $u=\tan{(\sqrt{\beta}p_{0_x})}$ was considered. Using the Stirling and asymptotic formulas \cite{abramowitz1968handbook}
\begin{align}
    \Gamma(\alpha)\sim& \sqrt{2\pi} e^{-\alpha}\alpha^{\alpha-1/2},\\
    \Gamma(n+2\alpha)\sim& \sqrt{2\pi}e^{-2\alpha}(2\alpha)^{2\alpha+n-1/2},
\end{align}
it is straightforward to verify that, in the limit \(\alpha \to \infty\), the term in brackets in \eqref{eq:Wave_funct_GUP_px_4} reduces to
\begin{equation}
    \lim_{\alpha\to\infty}\left[\frac{\sqrt{\beta}(n+\alpha)\Gamma^2(\alpha)\alpha^{n}}{2^{1-2\alpha}\pi\Gamma(n+2\alpha)n!}\right]^{1/2}=\frac{1}{\sqrt{\pi^{1/2}2^{n}n!}}.\label{eq:limit_N_const}
\end{equation}
On the other hand, by using the expansion \(\ln(\cos{x}) = -\frac{x^2}{2} - \frac{x^4}{12} - \cdots\) and the approximation \(\alpha \approx \frac{1}{\beta}\), we obtain
\begin{align}
    \lim_{\beta\to0}\cos^{n+\alpha}{(\sqrt{\beta}p_{0_x})}=& \lim_{\beta\to0}e^{n\ln{\cos{(\sqrt{\beta}p_{0_x})}}}e^{\alpha\ln{\cos{(\sqrt{\beta}p_{0_x})}}}\nonumber\\
    \sim& e^{-p_{0_x}^2/2}.\label{eq:Limit_cos}
\end{align}
In the limit $\alpha\to \infty$ (non-relativistic limit) the relativistic Hermite polynomial $H_{n}^{(\alpha )}$ turns into the Hermite polynomial $H_n(\xi)$ \cite{10.1063/1.530606}, and equivalently, its argument reduces to $\lim_{\beta\to0}\tan{(\sqrt{\beta}p_{0_x})}/\sqrt{\beta}=p_{x_0}$. So, we conclude that
\begin{equation}
    \lim_{\alpha\to\infty}H_{n}^{(\alpha )}(\sqrt{\alpha}\tan{(\sqrt{\beta}p_{0_x})})=H_{n}(p_{0_x}).\label{eq:Limit_H_relat_to_H}
\end{equation}
Combining this limit with those found in \eqref{eq:limit_N_const} and \eqref{eq:Limit_cos}, we find that the wave function \eqref{eq:Wave_funct_GUP_px_3}, derived using the minimal uncertainty approach, reduces to, as expected, the wave function \eqref{eq:Usual_Solut_1}.

\subsection{Fourier transform}
The wave function in~\eqref{eq:Wave_funct_GUP_px_3} is defined in momentum space, in terms of the auxiliary variable \( p_{0_x} \). In order to extract physical information from our results, it is necessary to express the wave function inside the black hole in terms of the variable \( x \), which is related to the radial coordinate, since the metric components in~\eqref{eq:sch-inter_a} depend on this variable. The validity of our results is restricted to the semiclassical limit \( R_s \gg \ell_{\mathrm{Pl}} \), which implies \( n \to \infty \). Therefore, we shall express the solution in~\eqref{eq:Wave_funct_GUP_px_3} in this limit. To this end, we first rewrite the Gegenbauer polynomials \( C_n^{(\alpha)} \) in terms of the Jacobi polynomials \( P_n^{(\alpha,\beta)} \)~\cite{abramowitz1968handbook}
\begin{equation}
    C_n^{(\alpha)}(x)=\frac{\Gamma(\alpha+\frac{1}{2})\Gamma(2\alpha+n)}{\Gamma(2\alpha)\Gamma(\alpha+n+\frac{1}{2})}P_n^{(\alpha-\frac{1}{2},\alpha-\frac{1}{2})}(x).\label{eq:Gengen_pol_in_Jacobi_poly}
\end{equation}
The asymptotic behavior of \( P_n^{(\alpha,\beta)}(x) \) as \( n \to \infty \) is given by the following approximation~\cite{szeg1939orthogonal}
\begin{multline}
    P_n^{(\alpha,\beta)}(\cos{\theta})=(n\pi)^{-1/2}\left(\sin{\frac{\theta}{2}}\right)^{-\alpha-\frac{1}{2}}\left(\cos{\frac{\theta}{2}}\right)^{-\beta-\frac{1}{2}}\\
    \times\cos{\left[\left(n+\frac{\alpha+\beta+1}{2}\right)\theta-\left(\alpha+\frac{1}{2}\right)\frac{\pi}{2}\right]}+\mathcal{O}(n^{-1/2}),
\end{multline}
which is valid uniformly over the interval \( 0 < \theta < \pi \). Substituting this approximation into \eqref{eq:Gengen_pol_in_Jacobi_poly} allows us to obtain the asymptotic expression of \( C_n^{(\alpha)} \) as \( n \to \infty \), given by
\begin{multline}
    C_n^{(\alpha)}(\sin{(\sqrt{\beta}p_{0_x}}))=\frac{2^{\alpha}(n\pi)^{-1/2}\Gamma(\alpha+\frac{1}{2})\Gamma(2\alpha+n)}{\Gamma(2\alpha)\Gamma(\alpha+n+\frac{1}{2})}\\
    \times\cos^{-\alpha}{(\sqrt{\beta}p_{0_x})}\cos{\left[(n+\alpha)\left(\frac{\pi}{2}-\sqrt{\beta}p_{0_x}\right)-\frac{\alpha\pi}{2}\right]}+\mathcal{O}(n^{-1/2}).
\end{multline}
To obtain this expression, we have considered \(\theta = \frac{\pi}{2} - \sqrt{\beta}\, p_{0_x}\). According to the validity range of \(\theta\), it follows that \(-\frac{\pi}{2} < \sqrt{\beta}\, p_{0_x} < \frac{\pi}{2}\), which coincides with the domain of \(p_{0_x}\) defined in \eqref{eq:Self_adjoint_representation}.

Thus, in this limit \(n \to \infty\), the wave function in~\eqref{eq:Wave_funct_GUP_px_3} reduces to a simple expression, given by
\begin{equation}
    \bar{U}_n(p_{0_x})\sim\cos \left[(\alpha+n) \left(\frac{\pi }{2}-\sqrt{\beta } p_{0_x}\right)-\frac{\alpha\pi}{2}\right]. \label{eq:Wave_funct_GUP_px_3_n_large}
\end{equation}
With this expression, we can compute the Fourier transform within the minimum uncertainty approach, which is modified in this framework (see Ref.~\cite{PhysRevD.107.126009}). In our case, the transform takes the form
\begin{equation}
    U_n(x)\sim\int_{-\frac{\pi}{2\sqrt{\beta}}}^{\frac{\pi}{2\sqrt{\beta}}} e^{ip_{0_x} x/\ell_{\mathrm{Pl}}}\cos \left[(\alpha+n) \left(\frac{\pi }{2}-\sqrt{\beta } p_{0_x}\right)-\frac{\alpha\pi}{2}\right]\ dp_{0_x},\label{eq:Fourie_trans_GUP}
\end{equation}
which, after performing the integration, one obtain the wave function in \(x\)-space, in the minimal uncertainty approach, which can be write as
\begin{align}
     U_n(x)\sim\frac{\ell_{\mathrm{Pl}} e^{-\frac{i \pi  x}{2 \ell_{\mathrm{Pl}}\sqrt{\beta }}}}{\ell_{\mathrm{Pl}}^2\beta (\alpha +n)^2-x^2}\left[-i x \cos \left(\frac{\pi  \alpha }{2}\right) \left((-1)^n-e^{\frac{i \pi  x}{\ell_{\mathrm{Pl}}\sqrt{\beta }}}\right)\right.\nonumber\\
    +\left.\ell_{\mathrm{Pl}}\sqrt{\beta } \sin \left(\frac{\pi  \alpha }{2}\right) (\alpha +n) \left((-1)^n+e^{\frac{i \pi  x}{\ell_{\mathrm{Pl}}\sqrt{\beta }}}\right)\right],
\end{align}
or, equivalently
\begin{align}
     \Psi_n^{\text{GUP}}(x)\sim\frac{\ell_{\mathrm{Pl}} e^{-\frac{i \pi  x}{2 \ell_{\mathrm{Pl}}\sqrt{\beta }}}}{(x+\frac{R_s}{2})(\ell_{\mathrm{Pl}}^2\beta (\alpha +n)^2-x^2)}\left[-i x \cos \left(\frac{\pi  \alpha }{2}\right) \left((-1)^n-e^{\frac{i \pi  x}{\ell_{\mathrm{Pl}}\sqrt{\beta }}}\right)\right.\nonumber\\
    +\left.\ell_{\mathrm{Pl}}\sqrt{\beta } \sin \left(\frac{\pi  \alpha }{2}\right) (\alpha +n) \left((-1)^n+e^{\frac{i \pi  x}{\ell_{\mathrm{Pl}}\sqrt{\beta }}}\right)\right].\label{eq:wave_funct_x_space_GUP_n_large}
\end{align}
If we compare this wave function, derived within the minimum uncertainty framework, with that obtained through a standard quantization procedure, \eqref{eq:Usual_Solut_x_large_n}, we see that it still preserves an oscillatory character, although of a different nature. We emphasize that this wave function, which describes the dynamics of a Schwarzschild black hole, was derived in the semiclassical limit \( R_s \gg \ell_{\mathrm{Pl}} \), thereby restricting its validity to large quantum states characterized by \( n \).

We now proceed to analyze the convergence of the wave function in \eqref{eq:wave_funct_x_space_GUP_n_large}. We first consider its behavior in the asymptotic limit \( x \to \infty \), where it can be approximated as
\begin{equation}
    \Psi_n^{\text{GUP}}(x)\sim\frac{i\,\ell_{\mathrm{Pl}}\cos\!\big(\tfrac{\pi\alpha}{2}\big)}{x^{2}}\,
e^{-\frac{i \pi  x}{2 \ell_{\mathrm{Pl}}\sqrt{\beta }}}\,\left((-1)^n-e^{\frac{i \pi  x}{\ell_{\mathrm{Pl}}\sqrt{\beta }}}\right)\;+\;O\!\Big(\frac{1}{x^{3}}\Big), \qquad x\to\infty.
\end{equation}
In particular, we observe that \(\Psi_n^{\text{GUP}}(x) \to 0\) as \(x^{-2}\), indicating a decay with an oscillatory modulation of the wave function. This implies that, as in the standard case, an observer outside the black hole would perceive it as static, as expected.

On the other hand, by inspecting the wave function~\eqref{eq:wave_funct_x_space_GUP_n_large}, one might be tempted to conclude that it diverges at the points 
\(x = \pm \ell_{\mathrm{Pl}}\sqrt{\beta}(\alpha+n)\). However, upon closer analysis, it can be verified that these singularities are in fact removable. To demonstrate this, let us define \(x_{+} = \ell_{\mathrm{Pl}}\sqrt{\beta}(\alpha+n)\) and \(x_{-} = -\ell_{\mathrm{Pl}}\sqrt{\beta}(\alpha+n)\). The wave function~\eqref{eq:wave_funct_x_space_GUP_n_large}, in these regions, reduces to
\begin{align}
     \Psi_n^{\text{GUP}}(x_{\pm})\sim\pm\frac{2\ell_{\mathrm{Pl}} e^{\mp\frac{i \pi  n}{2}}\cos \left(\frac{\pi  \alpha }{2}\right)\sin \left(\frac{\pi  \alpha }{2}\right)(-1)^n(x_{\pm}\mp\ell_{\mathrm{Pl}}\sqrt{\beta}(\alpha+n))}{(x_{\pm}+\frac{R_s}{2})(x_{\pm}-\ell_{\mathrm{Pl}}\sqrt{\beta}(\alpha+n))(x_{\pm}+\ell_{\mathrm{Pl}}\sqrt{\beta}(\alpha+n))}.\label{eq:wave_funct_x_+_point_n_large}
\end{align}
From this expression, it becomes clear that the terms in the numerator cancels with the ones in the denominator, leaving the wave function finite in these regions.

Furthermore, from~\eqref{eq:wave_funct_x_space_GUP_n_large} we observe that there exists another singularity at \(x_{0} = -R_{s}/2 \approx -\ell_{\mathrm{Pl}}\sqrt{\beta}\,n\). In this region, the wave function can be written as
\begin{equation}
    \Psi_n^{\text{GUP}}(x_0)\sim-\frac{2\ell_{\mathrm{Pl}}^2\sqrt{\beta}(\alpha+n)e^{\frac{i \pi  n}{2}}\sin{\left(\frac{\alpha\pi}{2}\right)}(-1)^n}{(x_0+\ell_{\mathrm{Pl}}\sqrt{\beta}n)(x_0^2-\ell_{\mathrm{Pl}}^2\beta(\alpha+n)^2)}.\label{eq:GUP_Wav_func_in_x_0_reg}
\end{equation}
From here, in order to remove the singularity in the region \(x_{0}\), we fix \(\alpha\) to be an even integer, that is, \(\alpha = 2m\) with \(m = 0,1,2,\cdots\). 
This choice ensures that the numerator in \eqref{eq:GUP_Wav_func_in_x_0_reg} vanishes. With this condition, the wave function in~\eqref{eq:wave_funct_x_space_GUP_n_large} now takes the form
\begin{equation}
  \Psi_n^{\text{GUP}}(x)\sim\frac{i\ell_{\mathrm{Pl}}(-1)^{m}xe^{-\frac{i \pi  x}{2 \ell_{\mathrm{Pl}}\sqrt{\beta }}}}{(x+\frac{R_s}{2})(x^2-\ell_{\mathrm{Pl}}^2\beta (\alpha +n)^2)}\left((-1)^n-e^{\frac{i \pi  x}{\ell_{\mathrm{Pl}}\sqrt{\beta }}}\right).\label{eq:GUP_Wav_funct_finite}
\end{equation}
We can also verify that if \(\alpha = 0\) (a particular case within the even integers), 
the factor \(x^{2} - l_{Pl}^{2}\beta(\alpha+n)^{2}\) in the denominator vanishes. 
Consequently, the denominator exhibits a second--order zero at \(x_{0}\) 
(arising from the product of two vanishing factors). 
This implies that the singularity in this region is not removed. 
To avoid this issue, we restrict the values of \(\alpha\), now regarded as a quantum parameter, 
to \(\alpha = 2m\) with \(m = 1,2,\ldots\). 
This requirement enforces the regularity of the wave function in the vicinity of \(x_{0}\) 
and may be interpreted as an additional quantization condition imposed on the parameter \(\alpha\).

As a consequence of this new condition, from \eqref{eq:alpha_vs_beta}, we can express the deformation parameter 
\(\beta\) in terms of the quantum number \(m\), by
\begin{equation}
    \beta=\pm\frac{1}{\sqrt{2m(2m-1)}}.\label{eq:beta_quant_in_m_terms}
\end{equation}
From this expression, two possible values for \(\beta\) can be chosen: one positive, \(\beta > 0\), and one negative, \(\beta < 0\). In order for the black hole area spectrum in \eqref{eq:Area_gup_correct_1_1}---and consequently the mass \(M\)---to remain positive, the positive sign in \eqref{eq:beta_quant_in_m_terms} must be selected, which implies that \(\beta\) will be positive. Another observation that can be made from this expression is that choosing \(m=0\), which corresponds to \(\alpha=0\), would lead to \(\beta \to \infty\). This is physically inadmissible, since it would indicate that the effects introduced by \(\beta\) become overwhelmingly strong at the semiclassical level under consideration. This provides an additional physical argument for excluding the case \(\alpha = 0\). Conversely, in the limit \(\alpha \to \infty\), which corresponds to large values of \(m\), we recover the case \(\beta \to 0\), namely the standard scenario discussed in subsection \ref{subsec:Limit beta to 0}. Thus, small values of \(m\) enhance the magnitude of the effects introduced by \(\beta\), 
while these effects gradually diminish for large values of \(m\). In other words, this new quantum parameter fixes the admissible values that \(\beta\) can take.

As we have seen so far, the wave function \eqref{eq:GUP_Wav_funct_finite} obtained within this approach remains finite throughout the entire domain of \(x\), in contrast to the one derived in the standard quantization scheme, equation \eqref{eq:Usual_Solut_x_large_n}. We now proceed to demonstrate that this wave function is square-integrable in the Hilbert space 
$L^2\!\left(\mathbb{R}, \,\left(x + \tfrac{R_s}{2}\right)^2 dx\right)$, in the semiclassical limit \(R_s \gg \ell_{\mathrm{Pl}}\). From equation \eqref{eq:GUP_Wav_funct_finite}, we observe that the squared norm takes the form
\begin{eqnarray}
|\Psi_n^{\text{GUP}}(x)|^2\sim\frac{2\ell_{\mathrm{Pl}}^2x^2}{(x+\frac{R_s}{2})^2(x^2-\ell_{\mathrm{Pl}}^2\beta (\alpha +n)^2)^2}\left[1-(-1)^n\cos\left(\frac{\pi x}{\ell_{\mathrm{Pl}}\sqrt{\beta}}\right)\right],\label{eq:Mod_Square_wav_func_Gup_case}
\end{eqnarray}
and, by integrating this expression over the entire semiclassical domain of \(x\), we obtain
\begin{equation}
    \int_{-\infty}^{\infty}|\Psi_n^{\text{GUP}}(x)|^2\, \left(x + \tfrac{R_s}{2}\right)^2 dx\approx\frac{\ell_{\mathrm{Pl}}\pi}{\sqrt{\beta}}.\label{eq:Normaliz_wave_func_GUP_case}
\end{equation}
This expression is obtained by taking into account that \(n\) and \(\alpha\) are positive integers, which implies 
\(\cos(n\pi) = (-1)^n\), \(\sin(n\pi) = 0\), \(\cos\big((\alpha+n)\pi\big) = (-1)^{\alpha+n}\), and \(\sin\big((\alpha+n)\pi\big) = 0\). 
Furthermore, we have used the relations between hyperbolic and trigonometric functions, namely 
\(\cosh(-ix) = \cos(x)\) and \(\sinh(-ix) = -i \sin(x)\).


The result \eqref{eq:Normaliz_wave_func_GUP_case}, although restricted to the semiclassical limit, shows that the black hole wave function derived within the minimal-uncertainty framework is square-integrable, ensuring a well-defined Hilbert space and enabling the computation of expectation values. The normalization condition \eqref{eq:Normaliz_wave_func_GUP_case} depends only on the Planck length \(\ell_{\mathrm{Pl}}\) and the deformation parameter \(\beta\). In the limit \(\beta \to 0\), this condition is lost, yielding a divergent integral with no physical meaning. By contrast, the wave function \eqref{eq:Usual_Solut_x_large_n} is not square-integrable.

\begin{figure}[htbp]
  \centering
  \includegraphics[width=0.7\textwidth]{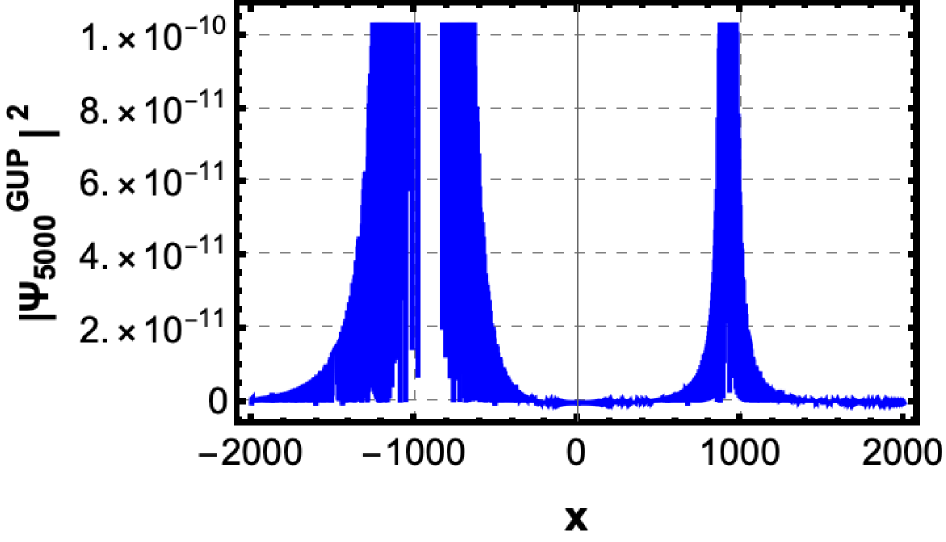} 
  \caption{Squared modulus of the wave function \eqref{eq:Mod_Square_wav_func_Gup_case}, obtained within the minimal-uncertainty formalism, shown in the semiclassical limit \(R_s \gg \ell_{\mathrm{Pl}}\).}
  \label{fig:Squar_wave_func_GUP}
\end{figure}

Figure \ref{fig:Squar_wave_func_GUP} shows the plot of the squared modulus of the wave function \eqref{eq:Mod_Square_wav_func_Gup_case} for a fixed state \(n=5000\) and quantum number \(m=15\), with the Planck length set to \(\ell_{\mathrm{Pl}}=1\). The same qualitative behavior is observed for other values of \(n\) and \(m\). These particular values are chosen merely for reference, with the purpose of illustrating the behavior of \eqref{eq:Mod_Square_wav_func_Gup_case}. In this plot, three distinct regions can be identified. The first extends from \(x \to \infty\) up to \(x_{+} = \ell_{\mathrm{Pl}}\sqrt{\beta}(\alpha+n)\). The second, an intermediate region, spans from \(x_{0_+} = \ell_{\mathrm{Pl}}\sqrt{\beta}n\), corresponding to the new black hole horizon, to \(x_{0_-} = -\ell_{\mathrm{Pl}}\sqrt{\beta}n\), where the classical singularity was located. Finally, the third region stretches from \(x_{-} = -\ell_{\mathrm{Pl}}\sqrt{\beta}(\alpha+n)\) to \(x \to -\infty\).

\begin{figure}[htbp]
  \centering
  \includegraphics[width=0.7\textwidth]{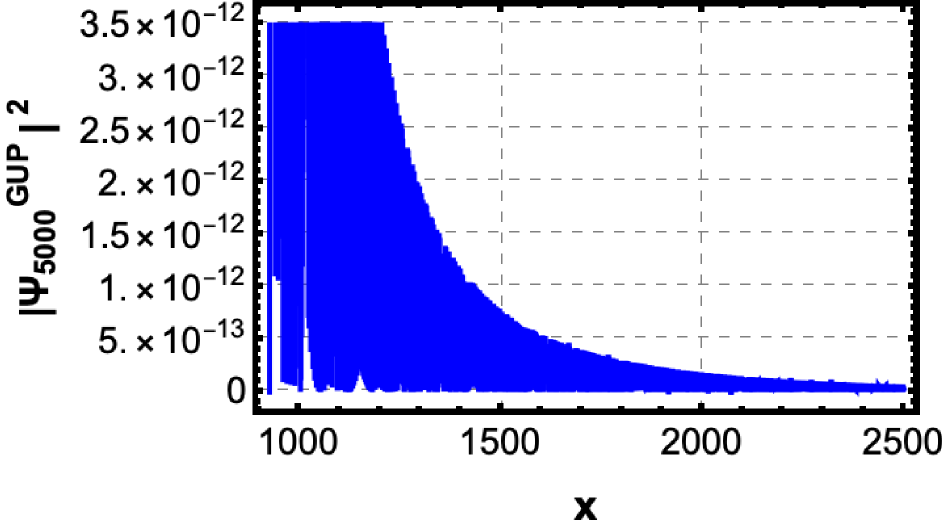} 
  \caption{Oscillation amplitude in the exterior region of the black hole, extending from \(x \to \infty\) down to the point \(x_{+}\).}
  \label{fig:Squar_wave_func_GUP_first_reg}
\end{figure}

\begin{figure}[htbp]
  \centering
  \includegraphics[width=0.65\textwidth]{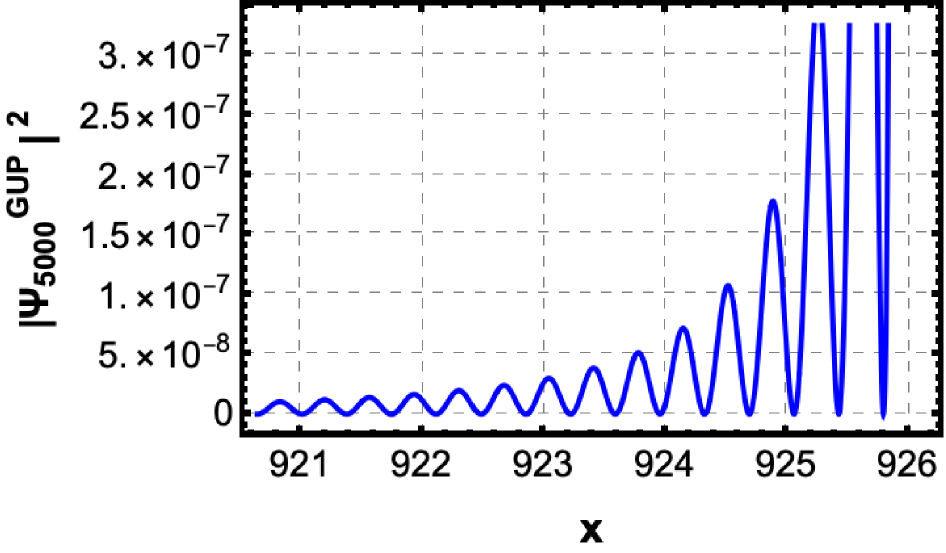} 
  \caption{Transition (tunneling) region across the event horizon, mediating the connection between the exterior and the interior of the black hole.}
  \label{fig:Squar_wave_func_GUP_transit_1_2}
\end{figure}

We may provide the following interpretation of these results. Let us begin with the first region, which lies outside the black hole. As can be seen (see Fig.~\ref{fig:Squar_wave_func_GUP_first_reg}), the oscillation amplitude $|\Psi_n^{\text{GUP}}(x)|^2$ vanishes in the limit \(x \to \infty\), where an external observer perceives the black hole as static. As one approaches the event horizon, the oscillation amplitude grows rapidly until it reaches the point \(x_{+}\), where it drops to zero. Beyond this point, a transition or tunneling, to interior of black hole, occurs into the second region, which begins at \(x_{0_{+}}\), as illustrated in Fig.~\ref{fig:Squar_wave_func_GUP_transit_1_2}.

\begin{figure}[htbp]
  \centering
  \includegraphics[width=0.65\textwidth]{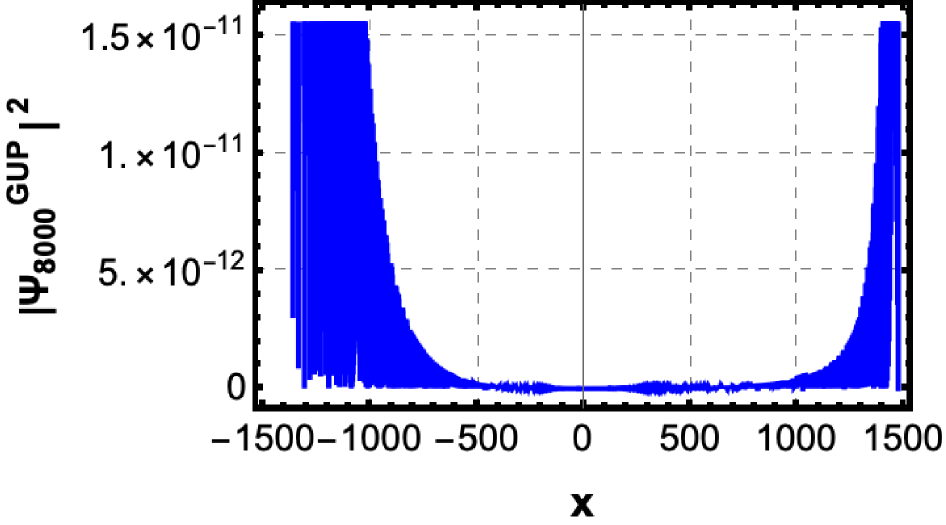} 
  \caption{Oscillation amplitude $|\Psi_n^{\text{GUP}}(x)|^2$ within the black hole interior, covering the region from $x_{0_+}$ to $x_{0_-}$, where the amplitude initially grows rapidly, then decreases, and finally rises again as it approaches $x_{0_-}$.}
  \label{fig:Squar_wave_func_GUP_int_BH}
\end{figure}
\begin{figure}[htbp]
  \centering
  \includegraphics[width=0.6\textwidth]{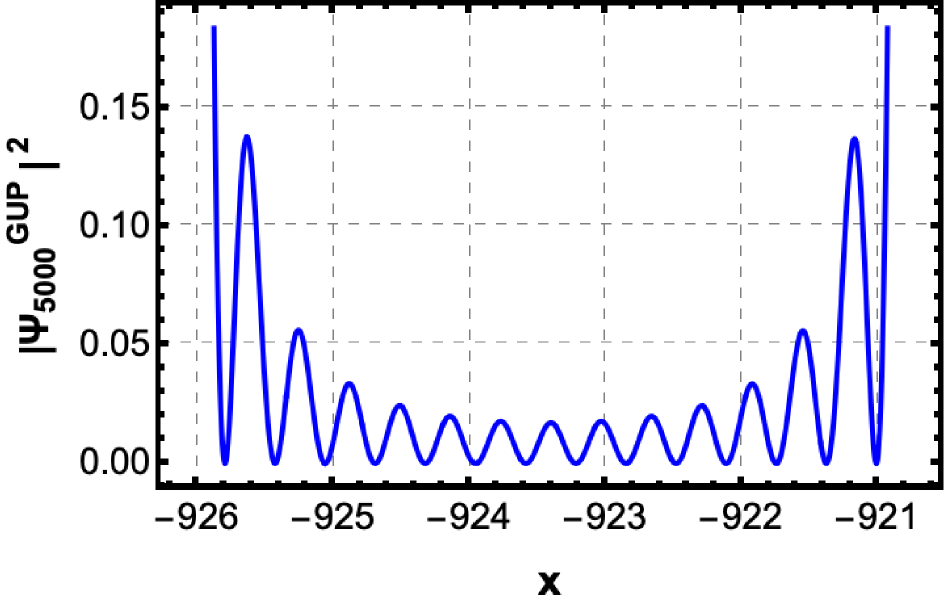} 
  \caption{Tunneling between the second and third regions, representing a quantum transition that connects the interior of the black hole with the exterior of a white hole.}
  \label{fig:Squar_wave_func_GUP_transit_2_3}
\end{figure}
\begin{figure}[htbp]
  \centering
  \includegraphics[width=0.7\textwidth]{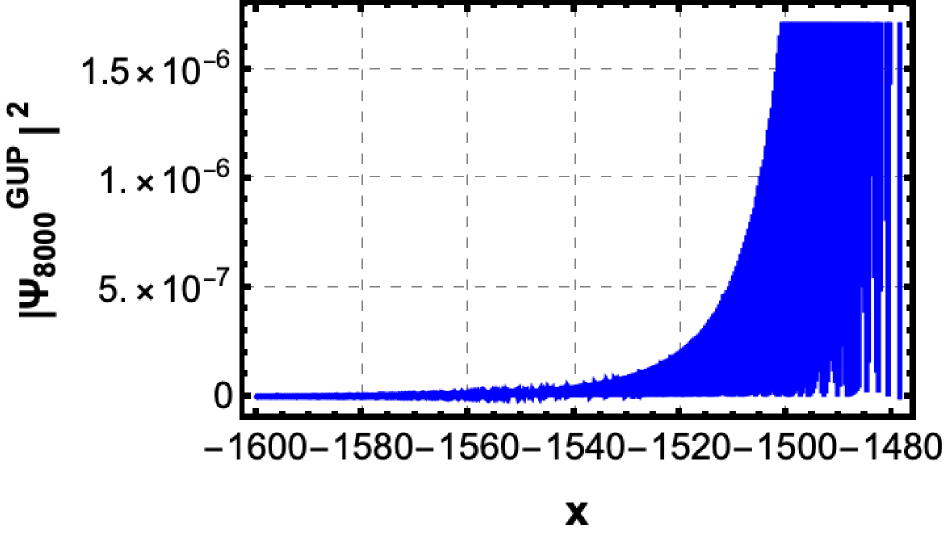} 
  \caption{Probability amplitude in the third region, which can be interpreted as the exterior of a white hole. The amplitude rises sharply near $x_{-}$ and gradually decays, vanishing in the limit $x \to -\infty$.}
  \label{fig:Squar_wave_func_GUP_ext_WH}
\end{figure}

Starting from the point $x_{0_+}$, the oscillation amplitude increases rapidly within this region. As one moves further inward into the black hole, the amplitude gradually decreases. However, approaching the vicinity of $x_{0_-}$, where $|\Psi_n^{\text{GUP}}(x)|^2$ vanishes, the oscillation amplitude rises again sharply (see Fig.~\ref{fig:Squar_wave_func_GUP_int_BH}). From $x_{0_-}$ up to $x_{-}$, a tunneling transition takes place toward the third region, as illustrated in Fig.~\ref{fig:Squar_wave_func_GUP_transit_2_3}. In this third region, the oscillation amplitude, as measured by $|\Psi_n^{\text{GUP}}(x)|^2$, grows rapidly near $x_{-}$, but diminishes progressively as one moves away from this point, eventually vanishing in the limit $x \to -\infty$ (see Fig.~\ref{fig:Squar_wave_func_GUP_ext_WH}).

Finally, by comparing the wave function obtained in the standard quantization scheme \eqref{eq:Usual_Solut_x_large_n} with that derived from the minimal-uncertainty approach \eqref{eq:GUP_Wav_funct_finite} in the semiclassical limit, we observe that they exhibit different behaviors both at the horizon and in the region where the black hole's physical singularity is located. For instance, in the standard case, the wave function is not finite throughout the domain of the variable \(x\) and, consequently, is not square-integrable in this domain. This prevents a proper definition of the Hilbert space to which it belongs and, therefore, does not allow the extraction of physical information. In contrast, due to the quantization based on the minimal-uncertainty approach, the wave function obtained is finite across the entire domain of \(x\) and is therefore square-integrable (see~\eqref{eq:Normaliz_wave_func_GUP_case}). Moreover, we find that the singularity previously residing inside the black hole is now replaced by a tunneling region that connects the black hole interior with the exterior of a white hole. This tunneling is of quantum nature, as its effect is determined by the quantum number \(m\) (and hence \(\alpha\)), which is directly related to the deformation parameter \(\beta\) through \eqref{eq:beta_quant_in_m_terms}; in the limit \(\beta \to 0\), this tunneling disappears.

\section{Discussion and conclusion}
\label{sec:Discussion and conclusion}
In this work, we explore the interior of a black hole from the perspective of two quantization approaches. We start with the eigenvalue equation \eqref{eq:Diff_equation_2}, originally derived by Mäkelä and Repo (1998) \cite{PhysRevD.57.4899}, which describes the dynamics inside the Schwarzschild black hole, setting the charge parameter \(Q = 0\). By rescaling the wave function and the radial coordinate, as shown in \eqref{eq:Change}, we reformulate this equation into the form of an eigenvalue equation for a quantum linear harmonic oscillator \eqref{eq:Diff_equation_4} \cite{obregon2001entropy}.

The first quantization scheme we employ is based on the standard approach, which in our case becomes feasible by considering the semiclassical limit \( R_s \gg \ell_{\mathrm{Pl}} \), or equivalently, by assuming large values of the black hole mass \( M \). This allows us to approximately define a Hilbert space of the form \( L^2\!\left(\mathbb{R}; (x + R_s/2)^2 dx\right) \). Within this framework, we find that the black hole's area spectrum \eqref{eq:Black_hol_Area_quantiz}, and consequently its radius \eqref{eq:Cuantiz_Schwar_radius}, are quantized by a discrete quantum number \( n \) and scale with the squared Planck length. In the asymptotic limit \( n \to \infty \), the area spectrum obtained in equation~\eqref{eq:Black_hol_Area_quantiz} reproduces the results predicted by other approaches~\cite{bekenstein1998quantum,PhysRevD.57.4899,PhysRevD.54.4982,obregon2001entropy}. Furthermore, the discreteness of the area spectrum implies that the mass \( M \) itself can only take discrete values, as expressed in equation~\eqref{eq:Espect_area_MR}. As a consequence, during Hawking evaporation, the black hole can only undergo transitions between mass eigenstates corresponding to these quantized levels.

The wave function given in equation~\eqref{eq:Usual_Solut_x_large_n} exhibits an oscillatory behavior that becomes increasingly pronounced for large black hole masses \( M \), or equivalently in the limit \( n \to \infty \), within the interior region of the black hole. As one moves away from the black hole, the amplitude of these oscillations gradually diminishes until it vanishes. Thus, for an observer at infinity, the black hole appears as a stationary object. These results are consistent with those reported in~\cite{PhysRevD.57.4899}. Moreover, the wave function displays a singular point at \( x = -\ell_{\mathrm{Pl}}\sqrt{2n} \) (see Figure~\ref{fig:Squar_wave_func_usuall}), around which strong oscillations with very large amplitude arise.

Also, the wave function derived in this quantization framework fails to be square integrable within the semiclassical domain under consideration, as shown in equation~\eqref{eq:Int_denom_K}. This lack of normalizability indicates that the state cannot correspond to a physically meaningful solution, thereby signaling an inconsistency of the standard quantization approach employed here.

The second quantization method we implement is based on the minimal uncertainty approach \eqref{eq:Modf_conmut_x_px}, in which the commutation relation between the dynamical variables is modified to introduce quantum gravity effects, quantified by the deformation parameter \(\beta\).  

Using this approach, we find that the black hole's area spectrum is modified, as shown in \eqref{eq:Area_gup_correct_2}, now scaling with the square of the quantum number \( n \). For large \( n \), both the area and the mass grow as \( A_s \sim 16\pi \beta n^2 \ell_{\mathrm{Pl}}^2 \), highlighting the increasing significance of quantum-gravitational effects at higher energy (mass) states. Unlike the linear growth found in equations~\eqref{eq:Black_hol_Area_quantiz} or~\eqref{eq:Espect_area_MR}, which depends solely on the Planck length, this quadratic dependence on \( n \) introduces a stronger sensitivity to the deformation parameter \(\beta\). Consequently, the black hole horizon expands more rapidly as it transitions to higher-mass states, while the mass \( M \) itself remains quantized, taking discrete values for each allowed quantum state. This behavior underscores the role of minimal-length effects in shaping the semiclassical structure of the black hole.

The wave function \eqref{eq:GUP_Wav_funct_finite}, which characterizes the quantum states of the modified black hole, remains finite throughout the entire semiclassical domain. To ensure the convergence of this wave function, we find that the deformation parameter \(\beta\) must depend on a discrete quantum parameter (see equation~\eqref{eq:beta_quant_in_m_terms}), denoted by \(m\), which restricts the allowed values it can take. To the best of our knowledge, this feature has not been reported in previous works.


Another noteworthy result is that the wave function obtained within this second quantization approach is square-integrable, as shown in equation~\eqref{eq:Normaliz_wave_func_GUP_case}. Consequently, it represents a well-defined physical state, with a normalization that depends solely on the quantum parameters \(\ell_{\mathrm{Pl}}\) and \(\beta\). In the limit \(\beta \to 0\), the normalization is lost, which allows us to interpret \(\beta\) as a parameter that effectively regularizes the wave function and ensures a physically meaningful state.


Finally, by graphically representing the squared modulus of the wave function, we find that the exterior and interior of the black hole are connected through a tunneling or transition region, characterized by a finite oscillatory probability amplitude. Additionally, we identify a tunneling region that connects the black hole interior with the exterior of a white hole. This tunneling is inherently quantum in nature, as its effect is determined by the quantum number \(m\), which is directly related to the deformation parameter \(\beta\) through equation~\eqref{eq:beta_quant_in_m_terms}; in the limit \(\beta \to 0\), this tunneling vanishes. A similar conclusion was reached in studies within the framework of Loop Quantum Gravity (LQG) \cite{Corichi_2016,PhysRevD.78.064040,MORALESTECOTL2021168401,PhysRevD.76.104030,https://doi.org/10.1155/2008/459290,PhysRevD.103.084038,Sobrinho_2023}.


We want to emphasize that the methods applied in this work and the results obtained by treating the black hole as a harmonic oscillator, both in standard quantization and in the minimal uncertainty approach, are novel in the literature. This opens up a potential area of study that will be explored in future projects.

\bmhead{Acknowledgements}

O. O and W. Y thanks to the grant by the University of Guanajuato CIIC 156/2024 ``Generalized Uncertainty Principle, Non-extensive Entropies, and General Relativity'', as well as the SECIHTI grant CBF2023-2024-2923 ``Implications of the Generalized Uncertainty Principle (GUP) in Quantum Cosmology, Gravitation, and its Connection with Non-extensive Entropies''. W. Y is supported by SECIHTI/Estancias Posdoctorales por México.

\bmhead{Author contribution} The authors contributed equally to this work.

\bmhead{Data availability}

No datasets were generated or analyzed during the current study.

\section*{Declarations}

\bmhead{Conflict of interest}

The authors declare no conflict of interest.



\bigskip


\bibliography{sn-bibliography}

\end{document}